# On the Encounter Rate of Open Star Clusters

A. D. Grinenko[1,2]   D. A. Kovaleva[2]

[1]*Moscow State University, Moscow, 119991 Russia*
[2]*Institute of Astronomy, Russian Academy of Sciences, Moscow, 119017 Russia*

Probable past and future close encounters of open clusters with known characteristics over 64 million years have been calculated by integrating the orbits of cluster centers in the Galactic potential using the galpy package. It has been shown that in the Galactic neighborhood of the Sun, pairwise cluster encounters at distances comparable to or smaller than their sizes occur at a characteristic rate of 35–40 events per 1 Myr. Close encounters between open clusters with a significant age difference occur at a rate of 15 events per Myr. It can be expected that in the Galaxy as a whole, such events occur an order of magnitude more frequently per unit time. Thus, dynamical interactions between stellar ensembles of different ages may not be too rare and could influence the properties of stellar populations. A pair of clusters with similar ages: HSC 1428 and Gulliver 22 was identified as a likely physically bound binary cluster system. A forecast of expected close encounters over the next 32 Myr has been provided for 490 pairs of clusters. Currently, 29 pairs of clusters are at their closest approach.

## 1. INTRODUCTION

The abundance of double and multiple stars in the field of the Galaxy is usually considered as a consequence of the process of formation of stars in groups (Duchêne and Kraus 2013).Thus, the components of binary star systems have a common origin, the same or very similar age, and initial abundance of chemical elements. The most common processes of binary star formation are considered to be the formation of bound systems during the fragmentation of a molecular cloud or during the fragmentation of a protostellar disk surrounding a formed star (see detailed discussion in Bate (2015), Tutukov and Cherepashchuk (2020)). It is also possible for binaries to form during inelastic collisions in the parent cluster, provided that excess kinetic energy can be released through interaction with a third body Cournoyer-Cloutier et al. (2021), tidal interactions (for very close pairs), or, most often, through a protostellar disk. Some wide pairs have been shown to form during the disintegration of open star clusters due to the low relative velocity of stars leaving the cluster relative to each other (Kouwenhoven et al. 2011, Rozner and Perets 2023, Tokovinin 2020). The components of pairs formed in such events may have an age difference determined by the duration of the star formation process in the cluster (up to several million years; see, for example, Jerabkova et al. (2019), Smilgys and Bonnell (2017)). However, there are known cases where the observational characteristics of the components of a binary system lead to significantly different estimates of the ages of the components (see discussion in Malkov and Kniazev (2022), Valle et al. (2016)). Such events are usually explained by inaccuracies in the definition of observational characteristics or evolutionary models. However, the interaction of a binary system with a single star, a binary star, or a high-multiplicity stellar system is theoretically possible during encounters outside the parent star cluster, including encounters between star clusters. Open star clusters are an important component of the Galaxy, and their evolution largely determines the characteristics of the stellar population (Krumholz et al. 2019, Lindegren and Dravins 2003). In particular, in clusters, a significant fraction of binary and multiple stars is formed (Bate 2015, Kouwenhoven et al. 2011, Rozner and Perets 2023). The publication of data from the Gaia space mission (Gaia Collaboration et al. 2016) has led to an abundance of new information on the number of open star clusters and moving groups of stars in the solar neighborhood, as well as to a significant increase in the accuracy of determining their characteristics. High-precision astrometric and photometric data from Gaia for more than 1.5 billion stars in our Galaxy have served as the basis for many projects using methods for automatically searching for regions of increased density in the 5D parameter space, including celestial coordinates, proper motions, and parallaxes, to search for clusters (see, e.g., Cantat-Gaudin and Anders (2020), Cantat-Gaudin et al. (2020, 2018), Castro-Ginard et al. (2020, 2021), He et al. (2021), Liu and Pang (2019), Sim et al. (2019) and others; a review of the used methods and the obtained results is given in Cantat-Gaudin and Casamiquela (2024)). The most complete homogeneous catalogue of star clusters discovered using automated search algorithms and machine learning methods applied to Gaia DR3 data (Gaia Collaboration et al. 2021, 2022), is the one authored by Hunt & Reffert (Hunt and Reffert 2024), and hereafter referred to as HR24.

New studies of the open star cluster population have revealed that, although some previously thought to exist clusters have turned out to be mirages (Cantat-Gaudin 2022), their number even at hundreds of parsecs from the Sun is significantly greater than previously thought, especially if we take into account stellar groups with common motion, which can be difficult to distinguish from clusters. Besides, the sizes of open clusters turned out to be significantly larger than previously thought. In this regard, in the Galaxy, unlike earlier ideas (de La Fuente Marcos and de La Fuente Marcos 2009, see, e.g., discussion in), a noticeable number of pairs of clusters that are located at a small physical distance from each other and interact gravitationally (e.g., Camargo (2021), Casado (2022), Kovaleva et al. (2020), Qin et al. (2023), Ye et al. (2022); see also reviews by Angelo et al. (2022), Vereshchagin et al. (2022)) were discovered. Typically, the clusters in these pairs are very similar in age and formed in the same giant molecular cloud. However, there are also other reliable examples of clusters and stellar groups of very different ages that are close to each other or approaching each other and have significant velocities relative to each other. Several examples of such encounters of different ages known to date can be listed: the approaching cluster Coma Ber and a moving stellar group (Fürnkranz et al. 2019, Sapozhnikov and Kovaleva 2021, Tang et al. 2019); the cluster NGC 1605 AB, which, as was stated by Camargo (2021), is the result of mixing of the populations of two clusters of different ages (this opinion has been opposed, though, by Anders et al. (2022)); the ongoing collision of the clusters IC 4665 and Collinder 350 (Piatti and Malhan 2022); etc.

If the populations of such clusters mix, it may be possible to form linked stellar systems whose components will have different ages. In this paper, we aim to answer the question of how frequent are the encounters of open star clusters, and what is the fraction of encounters between clusters of different ages. To do this, we construct a simplified model in which known clusters are considered as material points moving in the gravitational potential of the Galaxy, integrate their galactic orbits, and record the moments of approach. In Section 2, the data, approximations, and used methods are described; In Section 3, we present the obtained results, Section 4 is devoted to their discussion, and in Section 5, conclusions are provided.

## 2. DATA AND METHODS

We used data on clusters and their characteristics from the most complete modern homogeneous catalog of open star clusters (Hunt and Reffert 2024), HR24, which is the result of a systematic search for star clusters among the sources of the Gaia DR3 catalog and their comprehensive homogeneous study. The HR24 catalogue contains data for 7167 star clusters (including 134 globular clusters) at distances of up to 10 kpc from the Sun. In this case, radial velocities were determined for 5749 open star clusters, and we limit further consideration to them.

The following cluster parameters from HR24 were used to integrate the galactic orbits:

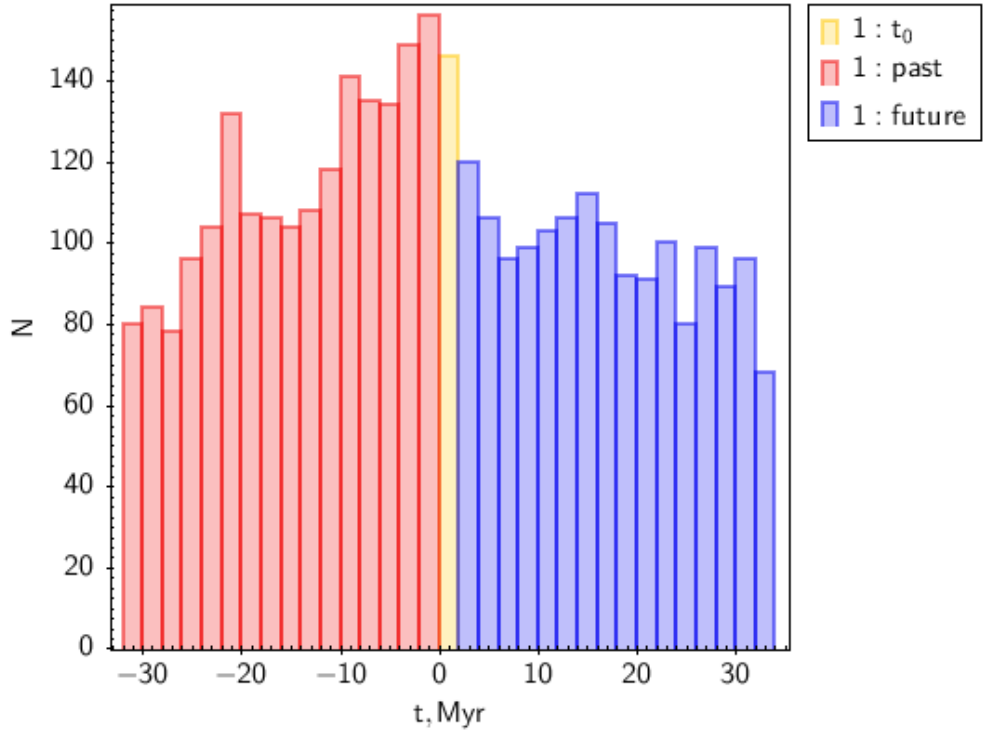

**Figure 1**. Distribution of the number of cluster approaches over time. The horizontal axis shows the moments of time from –32 to +32 million years relative to the present with a step of two million years. The vertical axis shows the number of cluster approaches. The number of encounters in the past is shown in red on the histogram, the number of encounters at the present moment $t = 0$ is shown in yellow, and the number of encounters in the future is shown in blue.

- coordinates RA_ICRS, DE_ICRS;

- parallaxes Plx;

- proper motions pmRA, pmDE;

- radial velocities RV.

  For subsequent analysis of the encounters, other characteristics of clusters from the HR24 catalogue were also used:

- cluster radii in parsecs: total *rtotpc*, Jacobi radius *rJtotpc*, core radius *rcpc*;

- logarithms of ages (median, first and third quartile): logAge16, logAge50, logAge84;

- total masses of clusters: MassTot.

*galpy* programming package (Bovy 2015) allows to integrate the orbits of objects in the gravitational potential of the Galaxy, provided that their three-dimensional coordinates and velocities at the present time are known. In the calculations, the three-component model of the galactic potential MWPotential2014 for our Galaxy was used. MWPotential2014 represents some simplification of the model Bovy and Rix (2013) and includes

- a bulge with a power density profile with an exponential cutoff, with the power degree is equal to –1.8, and the cutoff radius is 1.9 kpc;

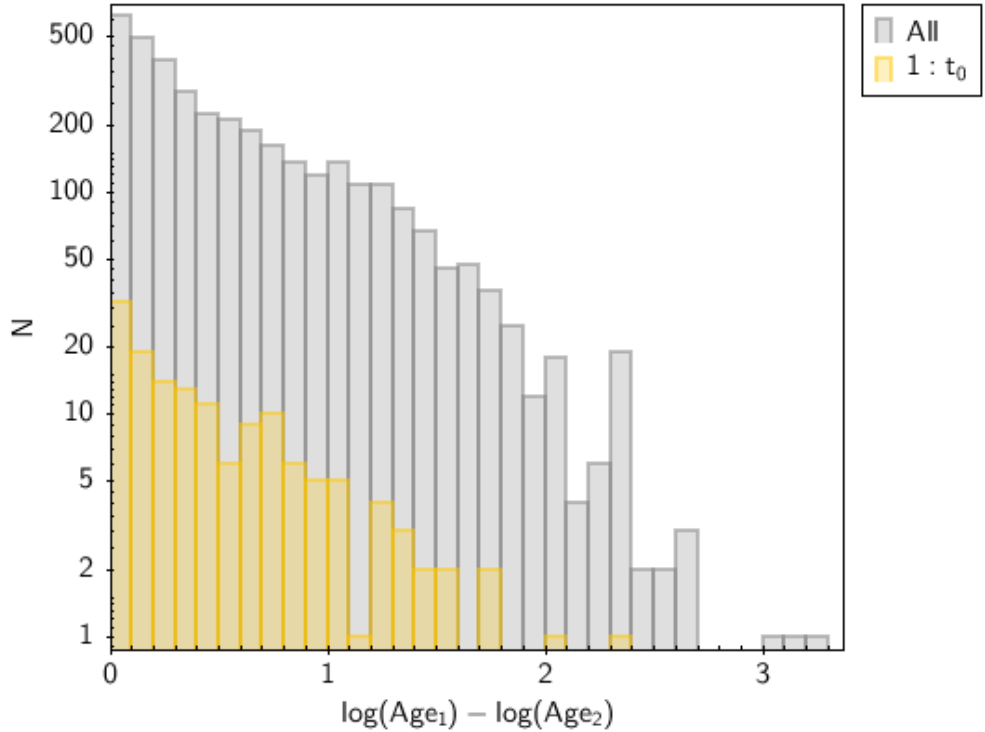

**Figure 2**. Distribution of approaches by the difference in ages of the approaching pair of clusters. The horizontal axis shows the differences in the logarithms of ages in pairs of clusters. The vertical axis shows the number of encounters between clusters with a given age difference. The total number of encounters is shown in grey on the histogram, and the number of encounters with a given age difference at the current moment in time is shown in yellow

- disk in the form of the Miyamoto–Nagai potential (Miyamoto and Nagai 1975);
- a dark matter halo described by the Navarro–Frank–White potential (Navarro et al. 1996).

The numerical parameters of these three components, the disk and halo size scales are selected to best match important observable characteristics of the Galactic population (Bovy 2015).

Orbits were integrated for each center of open clusters with determination of spatial-kinematic characteristics at the nodes of time scale with a step of 2 million years from the present moment to 32 million years into the past and into the future. In this way, positions were calculated for all 5749 open star clusters with known radial velocities. At each step, mutual distances were calculated for all pairs of clusters.

## 3. RESULTS

For the purposes of this study, a convergence of a pair of clusters was considered to be a position, in which the centers of the pair of clusters were located at a distance of no more than 20 pc from each other. The catalogue contains 146 pairs of clusters, the distance between the centers of which is currently ( ) is 20 pc or less. At distance from the present moment both into the past and into the future, the calculated number of approaches tends to decrease 1), which is associated with the blurring in space of the ensemble of open star clusters observed in the vicinity of the Sun today, and the replacement of some of them by other clusters not yet included in the catalogs of known ones.

We considered the values of significant age differences to be 0.72. The selection of this criterion was determined by the following considerations. In the HR24 catalogue, the distribution of age estimates for the same cluster is characterized by a median interquartile distance of 0.24 dex. Taking this value as an analogue of the standard deviation (SD) for a symmetric distribution, we select a deviation greater than three SDs as significant. To date ( ), 38 pairs of approaching clusters out of 146 have a significant age difference. The distribution of approaches by the difference in ages of clusters in the approaching pair is shown in Fig. 2.

The detected approaches are shown in Fig. 3 in coordinates: the distance between the centers of the clusters at the moment of approach–the sum of the total radii of the approaching clusters. All calculated encounters are shown in grey, with encounters currently observed highlighted in yellow ( ). To distinguish pairs located close to each other at formation from true encounters, in Fig. 4 the same events in the same coordinates are marked with a color that is darker the greater the difference in age for the two clusters participating in the encounter. The same pair may correspond to several points if the convergence is observed in more than one time node.

The total radii of the clusters $rtotpc$ (Fig. 3, 4 demonstrate the sum of total radii for the pair of clusters $rtotpc_1 + rtotpc_2$) are the estimates of the sizes of clusters from above with allowance for their possible coronas and tidal tails. On the other hand, for the same clusters, the median value of the ratio of the Jacobi radius $rJtotpc$ to total radius of the cluster $rtotpc$ is $rJtotpc/rtotpc = 0.58$, while the median value of this ratio for the radius of the core is $rcpc/rtotpc = 0.36$. A comparison of the radii of clusters given in the HR24 catalogue with their approach distances shows that in most cases, the sum of the total radii of clusters significantly (two or more times) exceeds the minimal distance between their centers during approach (Figs. 3, 4). Thus, most encounters to a distance of less than 20 pc between cluster centers result in mutual penetration of clusters and mixing of the cluster populations in space.

In the Appendix (Table 1), a forecast and characteristics of the closest approaches of clusters for the next 32 million years and a list of pairs of clusters that are currently in closest approach are given. The table contains identifiers of approaching clusters (Name 1, Name 2), the epoch of closest approach from 0 to 32 million years, minimal distance between cluster centers $Min\_dist$ in parsecs, full cluster radii $R_1$, $R_2$ in parsecs, the relative velocity of the clusters at the moment of closest approach $V_{rel}$ in km/s, and the difference in the logarithms of ages (in years) $dlogAge$. Each pair of clusters appears in the list at most once with closest approach characteristics.

## 4. DISCUSSION

The relative velocity of clusters as they approach each other affects the outcome of their dynamic interaction and the probability of population mixing. The distribution of approach events by the relative velocity of clusters in a pair at a minimal distance between their centers is shown in Fig. 5. Most encounters, both in general and at present, occur with a relative velocity of clusters not exceeding 10–20 km/s (the median approach velocity is 12.9 km/s). Moreover, clusters with a significant difference in age and clusters of similar ages, as is expected, show different distributions. Clusters of similar ages have predominantly low velocities at the moment of approach (a narrow maximum is expressed at 1–3 km/s; for high relative velocities, the frequency of events decreases). Pairs of clusters with a significant age difference show a distribution of relative approach velocities that is close to uniform in the region from two to approximately 25 km/s, with a slight broad maximum in the region of relative velocities of 10–12 km/s.

The approaches of pairs of clusters on a diagram distance between centers–relative velocity of movement, with color marking the difference in the ages of the clusters are shown in Fig. 6. It is clearly seen that the smallest differences in velocities are predominantly, but not exclusively, observed among pairs of similar ages. However, among pairs of clusters with a significant age difference, there are those that are closely approaching each other and those that are approaching each other at a low relative velocity. The distributions of approaches by specific orbital energy of a pair of clusters without and with the distribution of differences in the logarithms of the ages of the components in the pair are shown in Figs. 7, 8. The specific orbital energy (related to the reduced mass) is calculated with no allowance for the size of the clusters and the external galactic field using the formula from [45]: (Vallado 2007):

$$E = \frac{|v_1 - v_2|^2}{2} - \frac{GM}{S},$$

where

$$M = M_1 + M_2$$

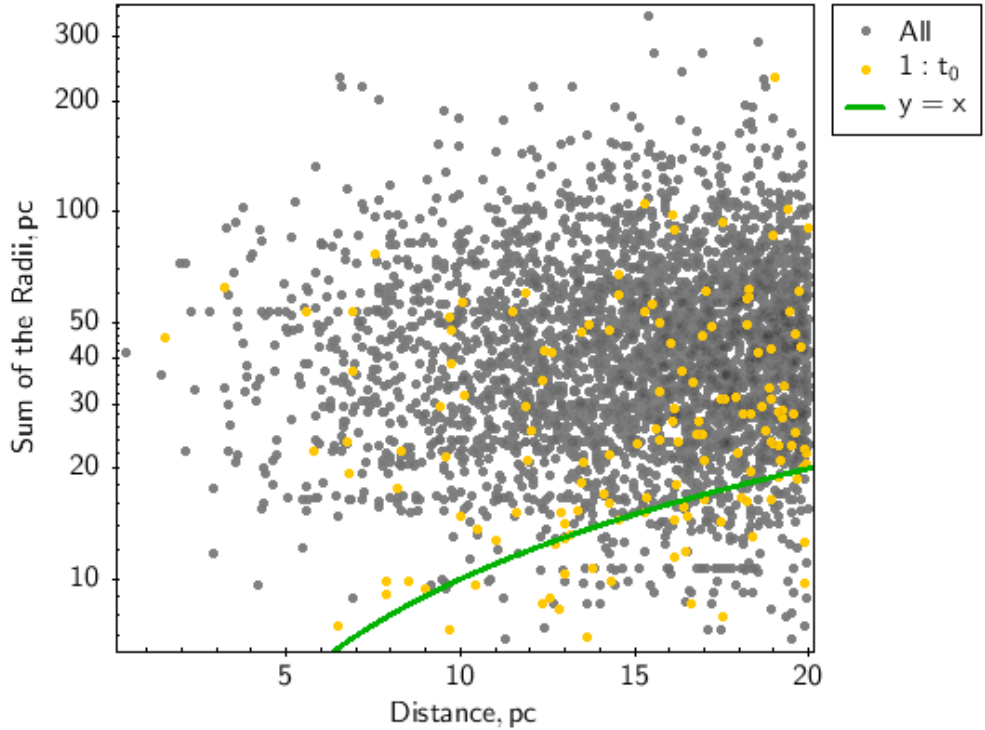

**Figure 3**. Comparison of the sizes of clusters and the distances between their centers. The horizontal axis shows the distances in parsecs between the centers of the clusters in the pair as they approach each other. On the vertical axis, there are shown the total radii of pairs of clusters $r_1 + r_2$ in parsecs. The grey color on the graph shows all the obtained approaches, and the yellow color shows the approaches of the clusters at the current moment in time. The green line corresponds to the condition of $y = x$, in which the total radii of two clusters when approaching each other are equal to the distances between the centers of these clusters.

– sum of masses of the clusters in the pair, S – distance between centers of the clusters.

No pairs with negative orbital energy were detected (except for one, which will be mentioned below), so in the used approximation, it is impossible to consider the formation of a gravitationally bound system when the clusters approach each other. However, if we consider a model of approach that takes into account the dynamic interactions of stars in clusters, the pattern may turn out to be significantly more complex. Excess kinetic energy can be spent either on increasing the velocities of cluster members (which can lead to the destruction of clusters or the ejection of individual stars) or on the destruction of binary and multiple systems in clusters. Among the pairs of clusters, there were found two that had an age difference of with $dlogAge > 1$ and relatively low specific orbital energy $E$. These are pairs: HSC 92 and HSC 63; SAI 122 and HSC 2695.

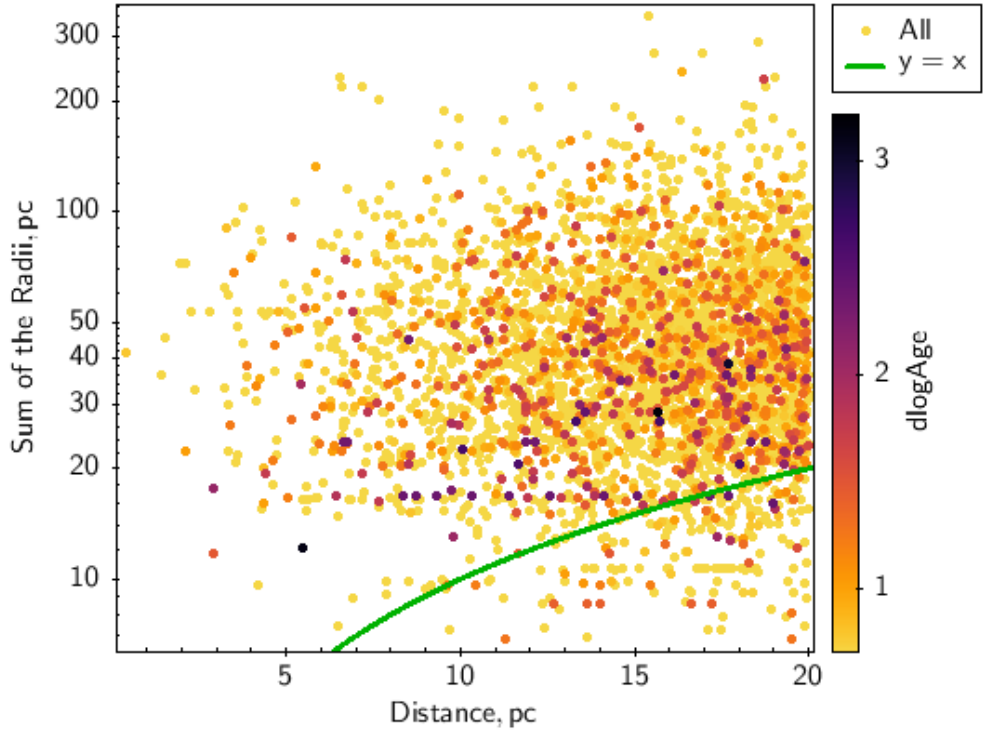

**Figure 4**. Comparison of the sizes of clusters and the distances between their centers with allowance for the distribution of differences in the ages of the components. The axes are similar to those in 3. The line $y = x$, in which the total radii of two clusters when approaching each other are equal to the distances between the centers of these clusters, is marked in green. The color indicates the difference in the logarithms of the ages of clusters in the approaching pair of clusters according to the legend on the right.

Among the approaches, only one pair of clusters was discovered, which in some epochs had a negative specific orbital energy of the system $E$, and is presently in its closest approach (the distance between the centers of the clusters is 5.6 pc): the clusters HSC 1428 and Gulliver 22. However, the difference in the logarithms of ages for this pair of clusters is only 0.107, and cannot be considered significant. Most likely, these are clusters that formed simultaneously close to each other in a single giant molecular cloud, which probably represent a gravitationally bound system.

We present (in the Appendix) a forecast of expected close encounters for 32 million years ahead, involving 490 pairs of clusters. For each pair, the characteristics are indicated only at the moment of closest approach. At the same time, 29 pairs of clusters are in closest proximity at present. Only for the components of 224 pairs, the approach is the only one during the considered period; in the remaining 266 pairs of clusters, one or both components participate in 1–4 more approaches during the considered period. The obtained estimate of the probable encounter frequency of open star clusters is 15 events per 1 million years, if for each pair each encounter is recorded only once, i.e., over 1 million years, 30 clusters participate in an encounter. This means, given the number of clusters in our sample, that on average for one cluster, a close approach can be expected once every 200 million years. This estimate is very close to the estimate made earlier in Fürnkranz et al. (2019) using a model with an isotropic distribution of 50000 clusters in the disk of the Galaxy with a radius of 30 kpc and a velocity dispersion corresponding to the velocity dispersion of the disk stars (one approach per cluster per 250 million years).

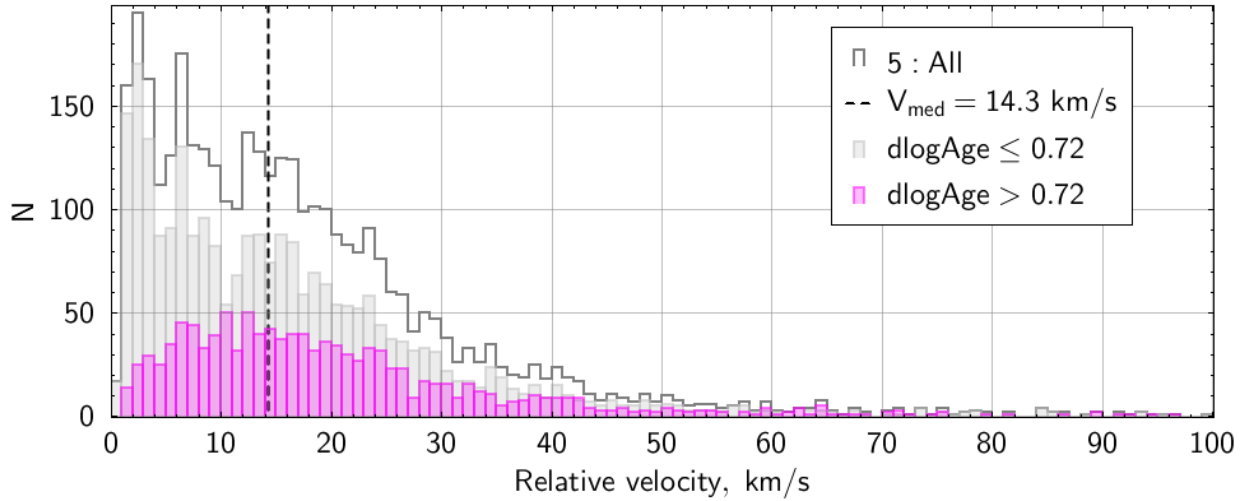

**Figure 5**. Relative velocities of approaching clusters. The horizontal axis shows the relative velocities of the clusters (in km/s) during their approach (the dependence is cut off at 100 km/s; the figure does not show the few approaches with relative velocities from 100 to 220 km/s). The vertical axis shows the number of approaches of clusters with given velocities. The total number of approaches is indicated by the grey outline. The dashed vertical line marks the median value of the relative velocity of the clusters at closest approach. The light gray histogram shows the convergence of pairs of clusters with an age difference less than the selected significance threshold; the purple histogram shows the convergence of pairs of clusters with a significant age difference.

In reality, the distribution of open star clusters in the Galaxy apparently obeys a more complex law than the isotropic distribution (Just et al. 2023, Scheepmaker et al. 2009), but the estimate of the total number of open star clusters may be close to that used Fürnkranz et al. (2019). Thus, based on the number of young massive clusters, based on the number of young massive clusters, 20,000 as the lower limit of the total number of clusters in the Galaxy are cited in Just et al. (2023); based on the estimate of the region of completeness of the sample of clusters [46], under the assumption of exponentially decreasing density, an estimate of 60,000–75,000 for the total number of clusters within a radius of 30 kpc from the center of the Galaxy can be obtained. Thus, the overall frequency of cluster encounters in the Galaxy must be at least an order of magnitude higher (150 new encounters per 1 million years).

## 5. CONCLUSIONS

The integration of the orbits of open star clusters from the HR24 catalogue that are located in the Galaxy mainly in the local neighborhood of the Sun has shown that paired approaches of clusters at distances comparable to their sizes and less occur in the galactic neighborhood of the Sun with a frequency of 35–40 events in 1 million years. At the same time, approaches of unrelated open clusters at distances comparable to their sizes and smaller occur in the galactic neighborhood of the Sun with a frequency of 15 events in 1 million years. The estimate of the total number of open star clusters in the Galaxy and their spatial distribution remains sufficiently uncertain, but it can be expected that the true frequency of cluster encounters in the Galaxy is at least an order of magnitude higher than the estimate for observed clusters.

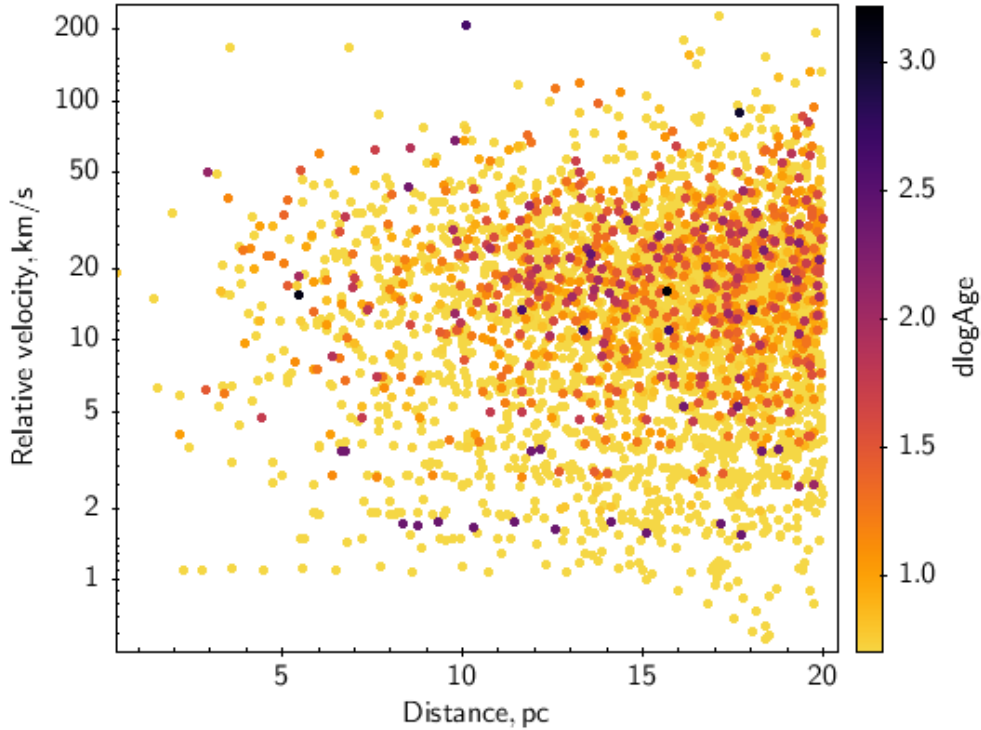

**Figure 6**. Relative velocity of clusters and the distance between their centers during approach, taking into account the difference in the logarithms of ages. The horizontal axis shows the distances in parsecs between the centers of the clusters as they approach each other. On the vertical axis are the relative velocities of clusters in km/s during approach. The color indicates the difference in the logarithms of the ages of clusters in the approaching pair of clusters according to the legend on the right.

A pair of clusters of similar age: clusters HSC 1428 and Gulliver 22, which represent a probable physical binary system of clusters, have been discovered. With the exception of this pair, in the used approximation (clusters as material points), no pairs with negative specific orbital energy at the moment of approach have been found. However, it can be assumed that when taking into account the dynamic interaction of cluster stars in a realistic model, hybrid clusters can form.

There are approaching pairs of clusters with large differences in the ages of their components. The formation of binary stars with components of different ages through the interaction of their populations is, therefore, not excluded. A detailed study of this issue requires dynamic modeling of the population of clusters during approach.

We provide in the Appendix a forecast of expected close encounters for 32 million years ahead, involving 490 pairs of clusters. At the same time, 29 pairs of clusters are in closest proximity at present.

The authors would like to thank the anonymous reviewer for insightful comments that helped improve the paper. The work uses data from the European Space Agency's (ESA) *Gaia* space mission (https://www.cosmos.esa.int/gaia), processed by the Gaia Data Processing and Analysis Consortium (DPAC, https://www.cosmos.esa.int/web/gaia/dpac/consortium). The interactive graphical visualizer and analyzer for tabular data TOPCAT (Taylor 2005) was used.

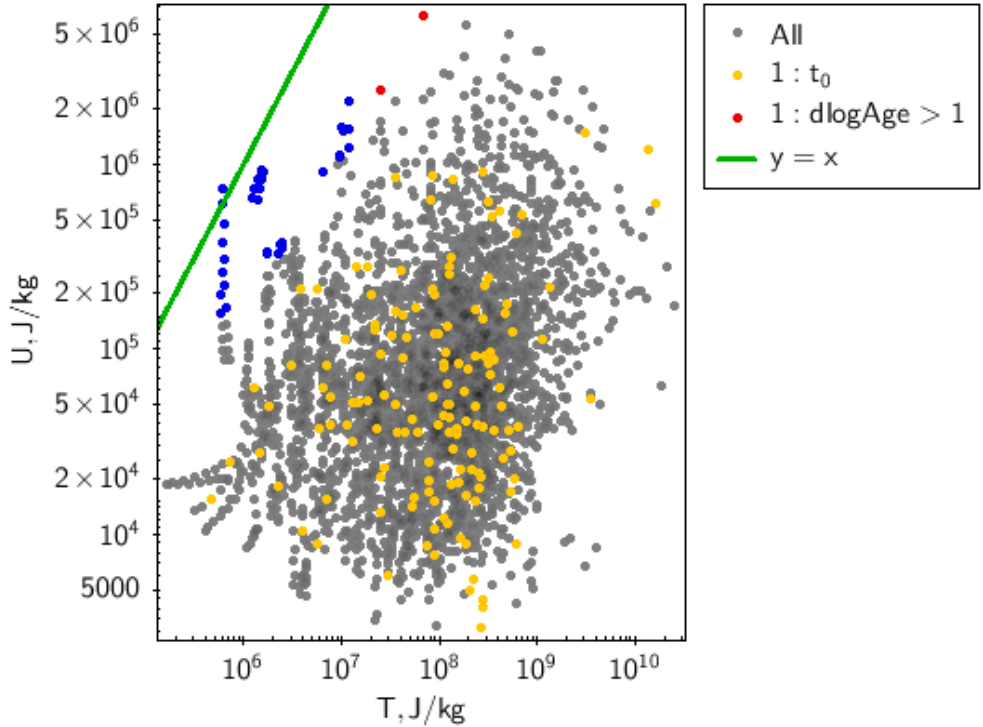

**Figure 7**. Distribution of encounters by specific orbital energy of pairs of clusters. The horizontal axis shows the specific kinetic energies of cluster approaches (in J/kg), and the vertical axis shows the specific potential energies of cluster approaches (in J/kg). The total number of encounters is shown in grey, the number of encounters at the present moment is shown in yellow, encounters for which the total specific energy of the pair of clusters is small are shown in blue, and encounters with a difference in the logarithms of the ages of the pair's components greater than 1 are shown in red. The green line shows the line of $y = x$, at which the specific potential and kinetic energies are equal.

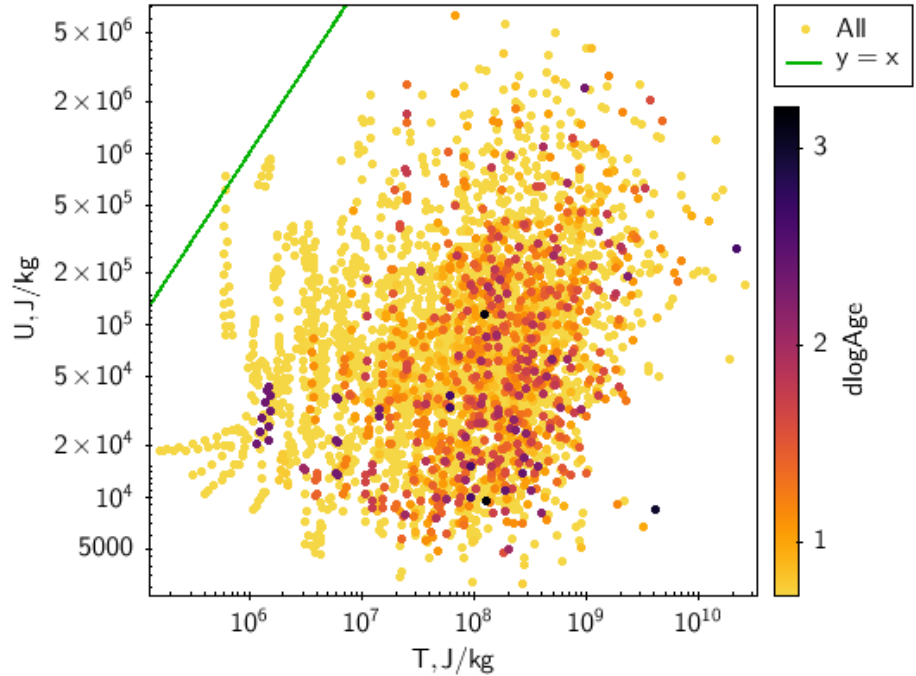

**Figure 8**. Distribution of approaches by specific energy of pairs of clusters with allowance for the distribution by differences in the ages of the components. The axes are similar to those in Fig. 7. The color indicates the difference in the logarithms of the ages of clusters in the approaching pair of clusters according to the legend on the right.

# APPENDIX

A list of pairs of open star clusters in the Galaxy that are currently in closest approach, as well as a forecast and characteristics of closest approaches of clusters for the next 32 million years with a step of two million years, is shown in the table below. The table contains identifiers of approaching clusters (Name 1, Name 2), the epoch of closest approach from 0 to 32 [million years] from the present time, minimum distance between cluster centers $Min\_dist$ in parsecs, total cluster radii $R_1$, $R_2$ and in parsecs, the relative velocity of the clusters at the moment of closest approach $V_{rel}$ in km/s, and the difference in the logarithms of ages (in years) $dlogAge$. Each pair of clusters is present in the list only once, at maximal proximity.

Table 1: Close encounters of open star clusters.

| Name 1 | Name 2 | t | Min_dist, пк | $R_1$, пк | $R_2$, пк | $V_{rel}$, км/с | dlogAge |
|---|---|---|---|---|---|---|---|
| UBC_1558 | UBC_550 | 0 | 19.9 | 12.0 | 10.6 | 34.8 | 0.84 |
| NGC_2311 | HSC_1723 | 0 | 16.2 | 8.5 | 5.9 | 30.5 | 1.01 |
| Melotte_22 | HSC_1318 | 0 | 19.8 | 35.6 | 7.2 | 24.3 | 1.34 |
| IC_2395 | HSC_2147 | 0 | 12.6 | 26.2 | 15.3 | 16.3 | 1.04 |
| UPK_535 | CWNU_1134 | 0 | 18.6 | 16.1 | 25.3 | 20.5 | 0.94 |
| HSC_883 | Aveni-Hunter_1 | 0 | 19.7 | 23.2 | 37.9 | 23.7 | 0.89 |
| UPK_194 | HSC_853 | 0 | 18.3 | 13.7 | 14.5 | 6.7 | 0.76 |
| HSC_2976 | HSC_103 | 0 | 19.5 | 10.7 | 12.5 | 23.1 | 0.78 |
| HSC_1991 | HSC_1977 | 0 | 19.0 | 15.7 | 7.5 | 23.4 | 0.75 |
| HSC_2468 | HSC_2394 | 0 | 12.4 | 24.3 | 17.4 | 14.8 | 1.04 |
| HSC_2085 | HSC_2082 | 0 | 11.9 | 15.3 | 14.3 | 25.2 | 1.7 |
| Theia_67 | OCSN_99 | 0 | 17.5 | 11.1 | 82.8 | 15.9 | 0.9 |
| NGC_2354 | CWNU_1064 | 0 | 6.9 | 35.5 | 17.9 | 16.8 | 1.8 |
| OCSN_100 | HSC_2919 | 0 | 14.3 | 5.1 | 4.8 | 9.3 | 0.88 |
| HSC_2976 | HSC_2971 | 0 | 9.7 | 10.7 | 41.0 | 18.3 | 0.98 |
| UBC_17a | OC_0339 | 0 | 14.3 | 11.2 | 4.9 | 17.5 | 1.23 |
| NGC_6193 | CWNU_1842 | 0 | 19.2 | 12.3 | 8.8 | 37.5 | 1.36 |
| HSC_674 | CWNU_1216 | 0 | 11.9 | 30.3 | 29.5 | 12.6 | 0.74 |
| Teutsch_182 | NGC_7429 | 0 | 18.1 | 17.3 | 11.0 | 20.8 | 1.06 |
| Theia_7 | HSC_1318 | 0 | 11.6 | 8.0 | 7.2 | 10.6 | 1.53 |
| HSC_1636 | CWNU_158 | 0 | 16.4 | 8.5 | 28.2 | 17.7 | 0.74 |
| HSC_1262 | HSC_1257 | 0 | 18.9 | 13.6 | 17.2 | 32.9 | 1.35 |

| | | | Table 1. Continued | | | | |
|---|---|---|---|---|---|---|---|
| Name 1 | Name 2 | t | Min_dist, пк | $R_1$, пк | $R_2$, пк | $V_{rel}$, км/с | dlogAge |
| HSC_570 | HSC_558 | 0 | 18.9 | 22.5 | 19.8 | 12.3 | 0.93 |
| HSC_1644 | CWNU_1015 | 0 | 13.7 | 10.9 | 38.6 | 22.6 | 1.56 |
| Theia_70 | HSC_95 | 0 | 19.4 | 15.7 | 85.4 | 19.4 | 1.29 |
| HSC_1667 | HSC_1644 | 0 | 16.8 | 13.8 | 10.9 | 20.1 | 2.04 |
| OC_0339 | ASCC_19 | 0 | 13.0 | 4.9 | 5.5 | 17.5 | 1.03 |
| HSC_1709 | ASCC_24 | 0 | 16.7 | 20.5 | 13.6 | 35.4 | 0.93 |
| OC_0128 | Bica_2 | 0 | 19.9 | 14.2 | 6.3 | 12.9 | 1.14 |
| Theia_1147 | Haffner_13 | 2 | 14.3 | 27.1 | 58.6 | 34.3 | 1.66 |
| HSC_480 | CWNU_1813 | 2 | 12.4 | 46.9 | 11.7 | 38.6 | 1.63 |
| Theia_1891 | HSC_2824 | 2 | 15.1 | 17.7 | 45.0 | 34.6 | 0.77 |
| UBC_338 | HSC_76 | 2 | 9.5 | 7.3 | 13.0 | 41.8 | 1.24 |
| Theia_3099 | Majaess_90 | 2 | 13.2 | 12.0 | 11.8 | 16.1 | 1.57 |
| Theia_65 | HSC_1347 | 2 | 17.6 | 18.8 | 42.7 | 9.2 | 1.42 |
| HSC_157 | HSC_103 | 2 | 10.7 | 34.8 | 12.5 | 25.3 | 1.69 |
| Theia_710 | NGC_7429 | 2 | 19.6 | 37.3 | 11.0 | 8.0 | 1.12 |
| Theia_3582 | HSC_1385 | 2 | 17.2 | 40.8 | 22.0 | 37.7 | 0.9 |
| UPK_567 | HSC_2449 | 2 | 11.2 | 46.5 | 47.5 | 28.7 | 0.85 |
| FSR_1421 | CWNU_182 | 2 | 6.2 | 20.5 | 30.1 | 17.9 | 1.19 |
| UBC_168 | OC_0181 | 2 | 17.4 | 37.4 | 4.4 | 30.6 | 1.6 |
| UBC_96 | ASCC_94 | 2 | 13.7 | 8.4 | 16.4 | 18.0 | 0.9 |
| Theia_710 | HSC_803 | 2 | 16.8 | 37.3 | 26.7 | 25.8 | 0.88 |
| UPK_198 | CWNU_1228 | 2 | 12.0 | 36.2 | 7.6 | 34.4 | 1.4 |
| OCSN_99 | HSC_2900 | 2 | 18.8 | 82.8 | 7.7 | 12.4 | 1.1 |
| UBC_541 | HSC_2754 | 2 | 15.0 | 10.1 | 7.6 | 42.6 | 0.93 |
| IC_348 | HSC_1257 | 2 | 10.9 | 21.5 | 17.2 | 24.3 | 1.28 |
| COIN-Gaia_18 | ASCC_14 | 2 | 17.4 | 8.3 | 16.2 | 18.6 | 0.74 |
| HSC_1134 | HSC_1127 | 2 | 15.7 | 8.9 | 14.3 | 16.6 | 1.03 |
| Theia_67 | HSC_157 | 2 | 18.8 | 11.1 | 34.8 | 15.2 | 1.24 |
| Platais_10 | HSC_2529 | 2 | 14.2 | 116.1 | 20.4 | 29.2 | 1.24 |

| | | | Table 1. Continued | | | | |
|---|---|---|---|---|---|---|---|
| Name 1 | Name 2 | t | Min_dist, пк | $R_1$, пк | $R_2$, пк | $V_{rel}$, км/с | dlogAge |
| Theia_65 | Melotte_25 | 2 | 6.7 | 18.8 | 55.1 | 32.8 | 1.79 |
| Collinder_106 | Basel_8 | 2 | 17.0 | 9.6 | 15.2 | 26.2 | 0.87 |
| OC_0339 | NGC_2068 | 2 | 19.3 | 4.9 | 12.4 | 23.4 | 1.13 |
| Theia_4 | HSC_892 | 2 | 11.2 | 10.3 | 8.3 | 17.8 | 1.72 |
| OCSN_91 | HSC_2816 | 2 | 17.7 | 4.9 | 12.1 | 1.5 | 2.39 |
| HSC_1630 | CWNU_1057 | 2 | 14.1 | 90.0 | 33.1 | 9.3 | 1.02 |
| OCSN_91 | HSC_2733 | 2 | 16.1 | 4.9 | 31.1 | 5.3 | 2.34 |
| OCSN_61 | OC_0339 | 2 | 19.1 | 10.0 | 4.9 | 4.9 | 0.83 |
| NGC_6883 | ASCC_110 | 2 | 19.3 | 15.0 | 25.5 | 47.8 | 1.66 |
| NGC_1333 | HSC_1257 | 2 | 19.7 | 4.5 | 17.2 | 32.8 | 1.33 |
| HSC_1495 | HSC_1437 | 2 | 14.8 | 19.2 | 15.8 | 9.3 | 1.11 |
| FSR_0551 | COIN-Gaia_34 | 4 | 8.0 | 14.4 | 15.3 | 16.4 | 1.21 |
| XDOCC_9 | Theia_38 | 4 | 11.0 | 18.3 | 49.0 | 15.4 | 1.4 |
| NGC_2451A | HSC_1746 | 4 | 13.7 | 17.6 | 79.2 | 23.8 | 0.84 |
| OCSN_92 | HSC_2784 | 4 | 14.0 | 18.9 | 28.0 | 27.0 | 1.35 |
| Theia_667 | CWNU_58 | 4 | 8.7 | 36.7 | 12.3 | 29.1 | 1.18 |
| Theia_230 | NGC_1662 | 4 | 13.7 | 39.3 | 93.1 | 25.3 | 1.32 |
| HSC_815 | FSR_0416 | 4 | 13.4 | 12.2 | 4.4 | 46.3 | 1.32 |
| HSC_892 | CWNU_1228 | 4 | 15.2 | 8.3 | 7.6 | 24.4 | 1.84 |
| Theia_72 | HSC_1849 | 4 | 14.7 | 31.2 | 43.9 | 15.4 | 1.2 |
| HSC_1989 | Alessi_3 | 4 | 6.6 | 38.6 | 20.5 | 28.3 | 1.53 |
| Theia_1797 | HSC_1590 | 4 | 10.8 | 33.9 | 23.5 | 11.2 | 1.08 |
| HSC_1977 | CWNU_1013 | 4 | 12.4 | 7.5 | 13.4 | 16.7 | 1.23 |
| Teutsch_145 | HSC_262 | 4 | 14.0 | 6.5 | 31.0 | 10.8 | 1.42 |
| NGC_743 | HSC_1049 | 4 | 18.3 | 21.7 | 24.5 | 14.6 | 1.02 |
| HSC_1714 | HSC_1640 | 4 | 14.7 | 26.6 | 38.9 | 12.5 | 0.99 |
| HSC_1644 | HSC_1640 | 4 | 19.9 | 10.9 | 38.9 | 16.9 | 1.88 |
| UBC_168 | Theia_4 | 4 | 19.1 | 37.4 | 10.3 | 29.0 | 1.42 |
| UBC_168 | HSC_862 | 4 | 16.0 | 37.4 | 8.2 | 33.9 | 1.32 |

| | | | Table 1. Continued | | | | |
|---|---|---|---|---|---|---|---|
| Name 1 | Name 2 | t | Min_dist, пк | $R_1$, пк | $R_2$, пк | $V_{rel}$, км/с | dlogAge |
| UBC_415 | CWNU_77 | 4 | 16.1 | 7.5 | 52.1 | 12.0 | 0.85 |
| HSC_2986 | HSC_2705 | 4 | 16.2 | 19.7 | 50.0 | 20.8 | 1.46 |
| UBC_168 | HSC_835 | 4 | 18.5 | 37.4 | 13.3 | 22.7 | 1.39 |
| UFMG_24 | HSC_2910 | 4 | 19.4 | 4.2 | 20.5 | 16.5 | 1.26 |
| UPK_198 | OC_0185 | 4 | 17.9 | 36.2 | 41.9 | 18.2 | 1.15 |
| UPK_198 | CWNU_1126 | 4 | 16.3 | 36.2 | 11.5 | 21.6 | 0.76 |
| UBC_1184 | Gulliver_19 | 4 | 10.4 | 17.3 | 32.0 | 57.4 | 0.95 |
| HSC_1733 | CWNU_1013 | 6 | 19.3 | 52.2 | 13.4 | 22.6 | 0.8 |
| UBC_97 | NGC_6561 | 6 | 15.7 | 7.9 | 6.1 | 18.8 | 0.85 |
| HSC_1470 | HSC_1460 | 6 | 5.9 | 17.2 | 6.1 | 7.6 | 1.27 |
| HSC_1463 | HSC_1361 | 6 | 18.3 | 33.3 | 18.6 | 20.3 | 2.05 |
| Theia_29 | HSC_2615 | 6 | 18.3 | 8.2 | 17.5 | 13.6 | 0.73 |
| Theia_2897 | FSR_0398 | 6 | 18.2 | 14.8 | 20.0 | 30.0 | 0.73 |
| OC_0227 | HSC_1037 | 6 | 19.7 | 12.9 | 9.7 | 33.7 | 1.4 |
| HSC_2244 | Gulliver_9 | 6 | 15.7 | 26.7 | 28.4 | 24.2 | 0.93 |
| Theia_19 | HSC_1067 | 6 | 9.8 | 5.5 | 44.0 | 25.9 | 1.77 |
| HSC_613 | Aveni-Hunter_1 | 6 | 8.0 | 19.3 | 37.9 | 40.9 | 1.32 |
| HSC_1390 | HSC_1318 | 6 | 18.0 | 13.4 | 7.2 | 13.5 | 2.6 |
| OC_0018 | HSC_87 | 6 | 12.6 | 5.8 | 9.3 | 112.7 | 1.26 |
| Theia_65 | HSC_1254 | 6 | 18.1 | 18.8 | 14.5 | 7.3 | 1.09 |
| Theia_558 | Collinder_95 | 6 | 16.5 | 60.5 | 11.9 | 13.2 | 1.67 |
| UPK_617 | HSC_2675 | 6 | 13.0 | 10.8 | 25.2 | 43.6 | 0.96 |
| HSC_2907 | HSC_157 | 6 | 17.6 | 23.7 | 34.8 | 13.0 | 1.32 |
| HSC_2931 | HSC_157 | 6 | 17.3 | 4.3 | 34.8 | 13.8 | 1.51 |
| FSR_0817 | CWNU_255 | 6 | 11.8 | 14.7 | 16.9 | 17.5 | 1.64 |
| HSC_1977 | HSC_1840 | 6 | 7.0 | 7.5 | 22.0 | 19.0 | 0.84 |
| HSC_1105 | CWNU_1029 | 6 | 11.4 | 44.3 | 9.3 | 19.1 | 1.52 |
| FSR_1744 | ESO_332-13 | 6 | 8.6 | 10.9 | 10.9 | 11.6 | 1.6 |
| HSC_1119 | HSC_704 | 6 | 7.6 | 8.5 | 21.7 | 62.3 | 1.7 |

| | | | Table 1. Continued | | | | |
|---|---|---|---|---|---|---|---|
| Name 1 | Name 2 | t | Min_dist, пк | $R_1$, пк | $R_2$, пк | $V_{rel}$, км/с | dlogAge |
| OCSN_98 | HSC_157 | 6 | 17.3 | 5.6 | 34.8 | 15.3 | 1.66 |
| HSC_2327 | HSC_2263 | 6 | 19.8 | 31.5 | 26.3 | 12.0 | 0.98 |
| HSC_157 | CWNU_1143 | 8 | 9.9 | 34.8 | 17.3 | 11.1 | 1.04 |
| HSC_447 | HSC_437 | 8 | 18.8 | 14.7 | 6.6 | 18.8 | 0.88 |
| HSC_1715 | HSC_1566 | 8 | 18.0 | 15.7 | 26.9 | 5.3 | 0.85 |
| HSC_450 | ASCC_107 | 8 | 16.7 | 18.4 | 14.1 | 6.6 | 0.96 |
| HSC_1874 | FSR_1129 | 8 | 18.6 | 28.0 | 17.5 | 59.9 | 1.21 |
| HSC_1961 | HSC_1920 | 8 | 15.2 | 47.4 | 40.1 | 5.8 | 0.77 |
| HSC_2106 | HSC_2071 | 8 | 12.5 | 23.0 | 10.3 | 8.6 | 0.94 |
| HSC_406 | CWNU_2024 | 8 | 17.9 | 18.3 | 7.5 | 24.4 | 1.36 |
| HSC_873 | HSC_725 | 8 | 19.8 | 8.7 | 13.9 | 19.5 | 1.54 |
| CWNU_2825 | CWNU_1841 | 8 | 18.8 | 8.0 | 10.6 | 33.5 | 1.13 |
| HSC_782 | HSC_594 | 8 | 19.8 | 20.8 | 31.0 | 11.6 | 0.83 |
| Theia_72 | HSC_2026 | 8 | 17.9 | 31.2 | 19.3 | 18.1 | 1.02 |
| Theia_4069 | FSR_0238 | 8 | 19.0 | 40.3 | 27.5 | 38.4 | 1.01 |
| Theia_5292 | Teutsch_175 | 8 | 12.6 | 24.6 | 7.1 | 24.4 | 0.89 |
| IC_4651 | HSC_2769 | 8 | 9.7 | 13.8 | 12.7 | 28.9 | 1.94 |
| UBC_415 | HSC_1052 | 8 | 17.7 | 7.5 | 7.9 | 4.7 | 0.83 |
| UBC_415 | CWNU_466 | 8 | 12.7 | 7.5 | 13.9 | 18.2 | 0.95 |
| Theia_117 | Teutsch_182 | 8 | 18.0 | 12.6 | 17.3 | 24.6 | 1.11 |
| Theia_178 | Teutsch_181 | 8 | 12.2 | 34.3 | 29.5 | 7.9 | 0.77 |
| Theia_11 | OCSN_24 | 8 | 16.3 | 10.4 | 18.0 | 27.1 | 1.01 |
| OCSN_88 | HSC_2068 | 8 | 16.3 | 224.5 | 15.8 | 11.5 | 0.91 |
| HSC_2653 | HSC_2641 | 8 | 16.4 | 34.8 | 23.4 | 14.2 | 0.88 |
| HSC_2622 | HSC_2612 | 8 | 12.2 | 8.4 | 26.0 | 18.1 | 0.74 |
| UPK_438 | OC_0343 | 8 | 17.1 | 15.9 | 20.0 | 29.6 | 1.02 |
| Theia_933 | HSC_1224 | 8 | 15.0 | 17.7 | 75.5 | 23.9 | 0.73 |
| HSC_180 | HSC_67 | 10 | 19.9 | 27.0 | 7.0 | 12.6 | 1.31 |
| Harvard_13 | CWNU_1538 | 10 | 18.8 | 27.6 | 6.8 | 48.3 | 1.03 |

Table 1. Continued

| Name 1 | Name 2 | t | Min_dist, пк | $R_1$, пк | $R_2$, пк | $V_{rel}$, км/с | dlogAge |
|---|---|---|---|---|---|---|---|
| Pozzo_1 | HSC_2452 | 10 | 19.2 | 12.1 | 51.8 | 21.6 | 1.22 |
| Stock_1 | OCSN_32 | 10 | 14.7 | 36.5 | 14.3 | 28.9 | 1.48 |
| UPK_189 | HSC_881 | 10 | 15.3 | 27.1 | 13.1 | 30.0 | 0.86 |
| Pismis_5 | HSC_2179 | 10 | 11.9 | 8.0 | 8.6 | 18.8 | 1.79 |
| HSC_2585 | Chamaleon_I | 10 | 12.0 | 63.2 | 11.6 | 14.0 | 1.06 |
| OC_0673 | NGC_6204 | 10 | 14.2 | 4.1 | 8.0 | 42.4 | 1.05 |
| UPK_116 | Stock_1 | 10 | 17.4 | 20.1 | 36.5 | 13.8 | 0.77 |
| HSC_1666 | HSC_1344 | 10 | 13.8 | 91.9 | 25.1 | 41.8 | 1.03 |
| Mamajek_1 | HSC_719 | 10 | 14.9 | 4.8 | 20.0 | 19.3 | 1.82 |
| BDSB_93 | ASCC_28 | 10 | 11.1 | 17.2 | 17.5 | 33.2 | 1.11 |
| Theia_26 | HSC_889 | 10 | 19.8 | 24.7 | 21.6 | 28.3 | 1.29 |
| HSC_2237 | HSC_1989 | 10 | 11.2 | 9.5 | 38.6 | 12.9 | 1.62 |
| HSC_1369 | HSC_1257 | 10 | 14.8 | 17.9 | 17.2 | 18.0 | 0.74 |
| Theia_69 | Platais_10 | 10 | 12.3 | 11.0 | 116.1 | 25.8 | 0.9 |
| HSC_823 | CWNU_1107 | 10 | 16.4 | 8.4 | 11.1 | 4.6 | 1.77 |
| Theia_622 | CWNU_236 | 10 | 16.3 | 22.5 | 12.9 | 8.9 | 0.95 |
| HSC_2662 | HSC_2630 | 10 | 13.9 | 25.1 | 25.0 | 9.1 | 1.11 |
| HSC_427 | HSC_252 | 10 | 17.9 | 17.0 | 38.3 | 45.7 | 1.13 |
| Teutsch_182 | CWNU_1216 | 10 | 14.3 | 17.3 | 29.5 | 19.0 | 1.42 |
| NGC_6167 | UBC_311 | 10 | 19.9 | 20.4 | 18.3 | 15.6 | 0.98 |
| Theia_1038 | HSC_255 | 10 | 19.0 | 28.3 | 25.4 | 23.5 | 0.86 |
| Theia_110 | HSC_2136 | 10 | 13.1 | 38.7 | 6.6 | 15.3 | 2.0 |
| HSC_1460 | HSC_865 | 10 | 4.2 | 6.1 | 29.1 | 20.0 | 0.96 |
| Stock_2 | HSC_1262 | 10 | 17.4 | 90.2 | 13.6 | 20.4 | 1.76 |
| IC_2602 | HSC_2464 | 10 | 15.9 | 35.6 | 11.7 | 10.6 | 0.96 |
| HSC_2871 | HSC_2816 | 12 | 10.8 | 18.7 | 12.1 | 13.7 | 1.62 |
| HSC_2712 | HSC_2690 | 12 | 18.4 | 25.9 | 3.8 | 13.3 | 1.19 |
| UPK_535 | Melotte_22 | 12 | 15.6 | 16.1 | 35.6 | 29.6 | 0.74 |
| HSC_2260 | HSC_2028 | 12 | 13.8 | 38.1 | 15.0 | 18.0 | 1.98 |

| | | | Table 1. Continued | | | | |
|---|---|---|---|---|---|---|---|
| Name 1 | Name 2 | t | Min_dist, пк | $R_1$, пк | $R_2$, пк | $V_{rel}$, км/с | dlogAge |
| HSC_725 | CWNU_1144 | 12 | 12.6 | 13.9 | 47.5 | 34.6 | 1.51 |
| Theia_436 | BH_164 | 12 | 19.8 | 19.2 | 131.2 | 16.8 | 0.77 |
| Theia_1858 | CWNU_128 | 12 | 19.0 | 18.4 | 21.9 | 38.3 | 1.01 |
| HSC_2597 | Alessi_9 | 12 | 17.5 | 63.0 | 31.4 | 9.1 | 1.06 |
| HSC_2712 | HSC_171 | 12 | 13.2 | 25.9 | 12.5 | 23.9 | 1.04 |
| HSC_2971 | CWNU_1205 | 12 | 14.8 | 41.0 | 24.7 | 22.2 | 1.07 |
| HSC_2662 | HSC_2529 | 12 | 14.8 | 25.1 | 20.4 | 8.6 | 1.13 |
| HSC_94 | ESO_522-05 | 12 | 11.7 | 7.1 | 4.7 | 6.2 | 1.49 |
| UPK_38 | Theia_622 | 12 | 14.6 | 12.9 | 22.5 | 8.3 | 1.48 |
| HXHWL_5 | HSC_1791 | 12 | 5.4 | 11.6 | 23.9 | 17.0 | 0.75 |
| NGC_6633 | Gaia_8 | 12 | 15.0 | 22.5 | 24.8 | 14.6 | 1.35 |
| UPK_629 | Theia_1643 | 12 | 16.5 | 18.1 | 23.6 | 23.6 | 1.7 |
| HSC_1481 | HSC_1447 | 12 | 15.4 | 57.5 | 95.1 | 13.3 | 0.73 |
| OCSN_49 | Delta_Cephei_Cluster | 12.0 | 9.1 | 86.1 | 35.8 | 20.3 | 0.84 |
| NGC_6334 | HSC_11 | 12 | 4.3 | 12.5 | 3.5 | 30.0 | 1.01 |
| Pozzo_1 | HSC_1746 | 12 | 15.6 | 12.1 | 79.2 | 21.7 | 1.28 |
| LISC-III_3668 | HSC_2174 | 12 | 9.3 | 54.2 | 11.5 | 19.6 | 0.86 |
| OCSN_70 | HSC_1707 | 12 | 17.8 | 3.4 | 20.1 | 18.1 | 1.28 |
| OCSN_70 | ASCC_24 | 12 | 13.1 | 3.4 | 13.6 | 20.1 | 0.72 |
| Theia_222 | Theia_70 | 12 | 15.4 | 21.2 | 15.7 | 10.9 | 0.95 |
| HSC_1942 | CWNU_1134 | 12 | 5.5 | 13.2 | 25.3 | 12.0 | 0.74 |
| OCSN_91 | HSC_2871 | 12 | 8.1 | 4.9 | 18.7 | 15.4 | 0.77 |
| Pismis_5 | NGC_2546 | 12 | 16.6 | 8.0 | 27.3 | 6.6 | 1.58 |
| OC_0497 | CWNU_1163 | 12 | 17.0 | 5.6 | 9.8 | 23.1 | 1.33 |
| HSC_1470 | CWNU_1178 | 12 | 19.3 | 17.2 | 27.6 | 9.8 | 1.08 |
| HSC_1871 | CWNU_1057 | 12 | 18.1 | 14.1 | 33.1 | 16.8 | 0.8 |
| Platais_12 | HSC_2529 | 12 | 16.9 | 46.2 | 20.4 | 20.3 | 1.31 |
| Platais_12 | HSC_2630 | 12 | 9.9 | 46.2 | 25.0 | 22.4 | 1.29 |
| OC_0017 | HSC_79 | 12 | 11.8 | 11.3 | 6.7 | 72.4 | 1.34 |

| | | | Table 1. Continued | | | | |
|---|---|---|---|---|---|---|---|
| Name 1 | Name 2 | t | Min_dist, пк | $R_1$, пк | $R_2$, пк | $V_{rel}$, км/с | dlogAge |
| HSC_1840 | FSR_1017 | 12 | 15.0 | 22.0 | 17.7 | 19.8 | 1.02 |
| NGC_6281 | HSC_2915 | 12 | 19.3 | 39.4 | 6.7 | 71.4 | 1.2 |
| Pismis_5 | HSC_2234 | 12 | 13.6 | 8.0 | 10.1 | 23.5 | 1.14 |
| Theia_169 | CWNU_1088 | 14 | 15.6 | 123.5 | 17.9 | 11.7 | 0.74 |
| IC_4665 | HSC_2705 | 14 | 8.4 | 23.7 | 50.0 | 25.6 | 0.8 |
| CWNU_58 | Alessi_37 | 14 | 16.2 | 12.3 | 63.6 | 18.7 | 1.18 |
| HSC_2687 | HSC_2120 | 14 | 12.8 | 16.3 | 53.1 | 28.1 | 0.73 |
| HSC_1986 | CWNU_1066 | 14 | 14.2 | 66.4 | 29.2 | 11.0 | 0.82 |
| Theia_1528 | ADS_16795 | 14 | 14.0 | 11.6 | 30.8 | 7.7 | 1.11 |
| Theia_5 | HSC_986 | 14 | 16.8 | 54.5 | 17.9 | 17.8 | 1.26 |
| Theia_29 | HSC_2609 | 14 | 16.0 | 8.2 | 23.3 | 12.1 | 0.77 |
| NGC_2215 | Collinder_132 | 14 | 16.5 | 13.3 | 31.4 | 33.0 | 1.51 |
| HSC_1523 | HSC_1314 | 14 | 16.4 | 10.3 | 73.6 | 10.5 | 0.95 |
| UPK_144 | Theia_713 | 14 | 16.7 | 7.0 | 94.3 | 20.5 | 1.1 |
| HSC_1119 | HSC_295 | 14 | 18.5 | 8.5 | 39.2 | 49.2 | 0.95 |
| Collinder_471 | ASCC_99 | 14 | 18.2 | 17.2 | 40.4 | 32.9 | 1.27 |
| HSC_2907 | HSC_188 | 14 | 15.5 | 23.7 | 12.7 | 7.1 | 1.65 |
| HSC_2239 | Gulliver_9 | 14 | 16.8 | 14.7 | 28.4 | 14.1 | 1.04 |
| HSC_1465 | HSC_782 | 14 | 15.5 | 22.9 | 20.8 | 27.2 | 2.29 |
| Theia_247 | HSC_238 | 14 | 12.2 | 70.2 | 16.4 | 17.5 | 0.99 |
| HSC_2983 | Collinder_394 | 14 | 14.9 | 6.7 | 22.3 | 24.4 | 0.78 |
| OCSN_98 | HSC_376 | 14 | 19.4 | 5.6 | 17.1 | 19.6 | 1.14 |
| OCSN_100 | HSC_376 | 14 | 15.6 | 5.1 | 17.1 | 19.9 | 1.23 |
| Ruprecht_53 | HSC_2092 | 14 | 17.1 | 14.6 | 10.9 | 20.5 | 1.15 |
| HSC_2610 | HSC_2068 | 14 | 17.0 | 43.3 | 15.8 | 10.6 | 1.48 |
| HSC_2645 | BH_164 | 14 | 15.2 | 38.5 | 131.2 | 20.0 | 1.52 |
| HSC_2641 | HSC_171 | 14 | 14.0 | 23.4 | 12.5 | 24.5 | 1.0 |
| Gulliver_9 | DBSB_64 | 14 | 19.5 | 28.4 | 10.9 | 36.4 | 1.01 |
| OC_0124 | Berkeley_90 | 14 | 16.6 | 10.3 | 27.7 | 64.3 | 0.93 |

Table 1. Continued

| Name 1 | Name 2 | t | Min_dist, пк | $R_1$, пк | $R_2$, пк | $V_{rel}$, км/с | dlogAge |
|---|---|---|---|---|---|---|---|
| OCSN_92 | HSC_2871 | 14 | 17.5 | 18.9 | 18.7 | 12.2 | 1.59 |
| Roslund_6 | Aveni-Hunter_1 | 14 | 15.5 | 25.2 | 37.9 | 16.4 | 1.17 |
| HSC_2452 | CWNU_1224 | 14 | 16.8 | 51.8 | 21.3 | 19.4 | 1.06 |
| HSC_94 | CWNU_448 | 14 | 11.5 | 7.1 | 16.1 | 26.5 | 1.07 |
| HSC_1894 | HSC_1843 | 16 | 16.9 | 15.6 | 27.8 | 15.1 | 0.92 |
| Theia_54 | HSC_1710 | 16 | 19.3 | 13.6 | 29.5 | 8.2 | 1.48 |
| OCSN_32 | ASCC_100 | 16 | 17.7 | 14.3 | 21.2 | 20.8 | 0.88 |
| HSC_2763 | HSC_2618 | 16 | 13.5 | 8.8 | 20.7 | 6.8 | 1.39 |
| HSC_2335 | HSC_1347 | 16 | 18.8 | 19.0 | 42.7 | 14.3 | 0.74 |
| HSC_2103 | HSC_2020 | 16 | 18.6 | 17.9 | 22.8 | 25.9 | 1.86 |
| Theia_366 | HSC_1894 | 16 | 16.7 | 19.6 | 15.6 | 12.9 | 0.91 |
| HSC_2314 | ASCC_85 | 16 | 19.6 | 34.3 | 41.4 | 59.6 | 1.68 |
| NGC_6800 | HSC_618 | 16 | 12.4 | 20.4 | 10.1 | 21.7 | 1.69 |
| HSC_2298 | Gulliver_9 | 16 | 15.2 | 24.1 | 28.4 | 12.0 | 1.0 |
| LK_10 | Bica_2 | 16 | 15.8 | 9.9 | 6.3 | 7.0 | 1.78 |
| IC_1396 | CWNU_460 | 16 | 15.1 | 14.7 | 16.7 | 24.1 | 1.43 |
| HSC_2303 | HSC_1648 | 16 | 16.5 | 70.9 | 24.9 | 20.8 | 0.87 |
| UPK_156 | BDSB_30 | 16 | 9.6 | 11.1 | 8.8 | 10.6 | 1.37 |
| OCSN_59 | CWNU_1054 | 16 | 8.7 | 15.0 | 18.9 | 6.8 | 1.31 |
| Theia_472 | HSC_2469 | 16 | 17.0 | 21.2 | 37.1 | 53.7 | 1.15 |
| Theia_181 | ASCC_79 | 16 | 16.5 | 41.0 | 15.8 | 16.3 | 1.31 |
| Alessi_37 | ASCC_125 | 16 | 17.3 | 63.6 | 4.2 | 16.9 | 1.53 |
| UBC_1105 | HSC_643 | 16 | 19.7 | 19.4 | 10.8 | 94.9 | 1.18 |
| Platais_12 | HSC_2636 | 16 | 14.2 | 46.2 | 19.0 | 19.6 | 1.18 |
| UPK_25 | HSC_911 | 16 | 16.8 | 17.7 | 20.1 | 39.4 | 0.97 |
| RSG_8 | HSC_956 | 16 | 17.4 | 21.7 | 23.0 | 16.9 | 0.87 |
| UFMG_66 | BH_211 | 16 | 19.9 | 11.8 | 9.2 | 19.2 | 1.1 |
| Teutsch_39 | NGC_7438 | 16 | 4.7 | 11.2 | 32.0 | 22.5 | 1.21 |
| HSC_415 | HSC_106 | 16 | 19.9 | 20.2 | 5.9 | 62.5 | 0.75 |

| | | | Table 1. Continued | | | | |
|---|---|---|---|---|---|---|---|
| Name 1 | Name 2 | t | Min_dist, пк | $R_1$, пк | $R_2$, пк | $V_{rel}$, км/с | dlogAge |
| UBC_534 | Harvard_13 | 16 | 17.2 | 19.7 | 27.6 | 41.3 | 0.91 |
| UBC_162 | HSC_779 | 16 | 16.8 | 15.6 | 13.7 | 12.5 | 0.99 |
| HSC_1146 | FSR_0686 | 16 | 15.9 | 6.4 | 10.0 | 15.5 | 1.09 |
| Theia_925 | ASCC_127 | 16 | 18.1 | 37.9 | 27.8 | 26.1 | 1.12 |
| RSG_8 | HSC_1125 | 16 | 15.4 | 21.7 | 17.4 | 24.9 | 1.03 |
| OCSN_92 | Mamajek_2 | 18 | 19.3 | 18.9 | 45.4 | 7.0 | 1.03 |
| HSC_1732 | HSC_1719 | 18 | 17.2 | 23.8 | 21.8 | 20.4 | 0.97 |
| Theia_171 | Teutsch_181 | 18 | 13.7 | 90.9 | 29.5 | 17.9 | 0.94 |
| UPK_82 | Teutsch_39 | 18 | 12.3 | 20.7 | 11.2 | 19.1 | 0.73 |
| UPK_172 | HSC_533 | 18 | 14.5 | 21.9 | 24.8 | 29.3 | 0.76 |
| HSC_1224 | ASCC_20 | 18 | 11.7 | 75.5 | 10.3 | 21.4 | 0.97 |
| UPK_169 | Theia_269 | 18 | 12.1 | 20.0 | 14.9 | 38.1 | 1.34 |
| HSC_1481 | HSC_1047 | 18 | 13.4 | 57.5 | 15.4 | 12.8 | 1.21 |
| HSC_2442 | CWNU_1265 | 18 | 11.9 | 32.3 | 10.6 | 7.8 | 1.65 |
| HSC_2630 | CWNU_513 | 18 | 16.3 | 25.0 | 13.8 | 10.2 | 1.41 |
| HSC_764 | CWNU_174 | 18 | 5.87 | 72.6 | 60.2 | 27.5 | 0.97 |
| UPK_640 | HSC_844 | 18 | 10.0 | 43.3 | 16.9 | 68.3 | 1.01 |
| Theia_118 | ASCC_24 | 18 | 17.5 | 19.1 | 13.6 | 5.0 | 1.77 |
| Theia_7 | HSC_2229 | 18 | 10.7 | 8.0 | 49.0 | 16.7 | 1.41 |
| Theia_101 | HSC_1465 | 18 | 18.0 | 31.5 | 22.9 | 34.0 | 1.49 |
| Theia_1232 | HSC_1469 | 18 | 16.2 | 14.5 | 31.1 | 21.4 | 1.13 |
| HSC_1765 | HSC_926 | 18 | 7.9 | 18.4 | 35.1 | 22.3 | 0.74 |
| HSC_1390 | HSC_926 | 18 | 19.3 | 13.4 | 35.1 | 15.7 | 1.01 |
| Theia_4310 | Theia_850 | 18 | 11.3 | 15.4 | 26.1 | 29.6 | 1.07 |
| UBC_159 | HSC_658 | 18 | 19.6 | 12.4 | 7.4 | 38.3 | 1.72 |
| NGC_7031 | CWNU_1276 | 18 | 9.2 | 14.6 | 15.7 | 54.9 | 1.03 |
| CWNU_1107 | CWNU_460 | 18 | 10.1 | 11.1 | 16.7 | 26.4 | 1.15 |
| HSC_440 | HSC_238 | 18 | 13.4 | 18.1 | 16.4 | 13.1 | 1.03 |
| HSC_2028 | ASCC_58 | 18 | 14.0 | 15.0 | 30.4 | 15.9 | 2.08 |

| | | | Table 1. Continued | | | | |
|---|---|---|---|---|---|---|---|
| Name 1 | Name 2 | t | Min_dist, пк | $R_1$, пк | $R_2$, пк | $V_{rel}$, км/с | dlogAge |
| Theia_227 | OCSN_12 | 18 | 18.8 | 39.5 | 29.8 | 17.4 | 0.77 |
| Teutsch_28 | Dolidze_1 | 20 | 14.6 | 15.3 | 12.7 | 45.7 | 0.84 |
| Teutsch_182 | HSC_2846 | 20 | 13.4 | 17.3 | 47.4 | 23.2 | 0.89 |
| NGC_5822 | CWNU_1173 | 20 | 19.9 | 25.6 | 45.3 | 22.0 | 1.02 |
| HSC_534 | Gaia_8 | 20 | 19.1 | 10.4 | 24.8 | 4.2 | 1.02 |
| UBC_1618 | HSC_2080 | 20 | 17.5 | 14.6 | 9.1 | 5.5 | 1.21 |
| HSC_2237 | HSC_2056 | 20 | 17.4 | 9.5 | 24.0 | 10.7 | 1.45 |
| HSC_1667 | ASCC_73 | 20 | 11.8 | 13.8 | 38.9 | 31.9 | 1.14 |
| HSC_2784 | HSC_1648 | 20 | 9.1 | 28.0 | 24.9 | 24.4 | 1.28 |
| Theia_4 | HSC_736 | 20 | 13.6 | 10.3 | 5.6 | 18.1 | 1.48 |
| UPK_113 | LDN_988e | 20 | 17.8 | 31.3 | 16.7 | 12.0 | 0.9 |
| NGC_2215 | HSC_2046 | 20 | 7.9 | 13.3 | 20.7 | 30.8 | 1.27 |
| HSC_2028 | CWNU_287 | 20 | 19.8 | 15.0 | 12.6 | 12.7 | 1.83 |
| HSC_884 | HSC_842 | 20 | 10.7 | 10.3 | 12.5 | 10.1 | 0.95 |
| UPK_438 | Mon_OB1-D | 20 | 12.1 | 15.9 | 3.8 | 16.1 | 1.75 |
| UPK_191 | Roslund_5 | 20 | 17.6 | 11.7 | 27.7 | 25.4 | 1.49 |
| Theia_1232 | Theia_229 | 20 | 10.3 | 14.5 | 62.2 | 30.3 | 0.95 |
| Theia_1256 | HSC_1958 | 20 | 11.9 | 11.7 | 13.6 | 21.8 | 1.65 |
| HSC_188 | CWNU_1004 | 20 | 6.4 | 12.7 | 4.2 | 8.6 | 1.88 |
| Theia_71 | Alessi_3 | 20 | 11.9 | 8.1 | 20.5 | 14.3 | 1.6 |
| Theia_1224 | OCSN_82 | 20 | 19.4 | 31.7 | 35.5 | 26.5 | 1.18 |
| Theia_759 | HSC_1920 | 20 | 19.9 | 26.9 | 40.1 | 14.5 | 1.15 |
| HSC_696 | FSR_0369 | 20 | 13.6 | 15.7 | 3.9 | 16.2 | 1.46 |
| HSC_1058 | Alessi_20 | 22 | 11.6 | 75.7 | 13.6 | 10.5 | 1.24 |
| Theia_865 | HSC_2587 | 22 | 16.2 | 62.7 | 10.6 | 8.8 | 0.8 |
| UBC_179 | HSC_576 | 22 | 18.9 | 35.7 | 15.6 | 34.4 | 1.21 |
| HSC_2983 | Alessi_80B | 22 | 16.5 | 6.7 | 14.1 | 28.3 | 1.24 |
| Teutsch_182 | HSC_629 | 22 | 17.7 | 17.3 | 18.7 | 13.0 | 1.38 |
| HSC_1213 | ASCC_21 | 22 | 19.9 | 31.8 | 10.6 | 20.1 | 1.5 |

Table 1. Continued

| Name 1 | Name 2 | t | Min_dist, пк | $R_1$, пк | $R_2$, пк | $V_{rel}$, км/с | dlogAge |
|---|---|---|---|---|---|---|---|
| HSC_1370 | HSC_1239 | 22 | 19.9 | 10.3 | 27.1 | 33.8 | 1.13 |
| NGC_2423 | HSC_2412 | 22 | 17.5 | 26.7 | 10.5 | 41.0 | 0.79 |
| OC_0018 | HSC_305 | 22 | 18.4 | 5.8 | 18.7 | 62.6 | 1.36 |
| NGC_6530 | CWNU_101 | 22 | 8.3 | 11.4 | 9.2 | 17.2 | 1.1 |
| NGC_7789 | HSC_877 | 22 | 19.1 | 28.4 | 19.2 | 46.6 | 0.87 |
| HSC_2648 | Alessi_20 | 22 | 20.0 | 9.3 | 13.6 | 32.3 | 1.63 |
| NGC_6735 | FSR_0165 | 22 | 19.5 | 27.9 | 11.1 | 53.9 | 0.8 |
| Majaess_90 | HSC_1974 | 22 | 18.7 | 11.8 | 26.8 | 6.8 | 1.06 |
| OC_0652 | HSC_381 | 22 | 12.6 | 9.7 | 37.9 | 37.2 | 0.83 |
| HSC_1648 | HSC_878 | 22 | 16.7 | 24.9 | 35.5 | 22.1 | 0.91 |
| HSC_1542 | ASCC_19 | 22 | 17.1 | 52.9 | 5.5 | 11.8 | 1.38 |
| HSC_1469 | HSC_1460 | 22 | 14.4 | 31.1 | 6.1 | 7.8 | 0.91 |
| UPK_118 | Czernik_2 | 22 | 16.1 | 15.9 | 21.8 | 60.2 | 0.93 |
| UPK_594 | Majaess_160 | 22 | 7.0 | 24.6 | 11.3 | 9.8 | 0.86 |
| Theia_227 | NGC_6568 | 22 | 18.3 | 39.5 | 7.9 | 26.6 | 1.24 |
| Theia_648 | ASCC_21 | 22 | 16.9 | 19.6 | 10.6 | 13.4 | 1.28 |
| UPK_540 | HSC_1724 | 22 | 13.6 | 26.4 | 21.7 | 15.3 | 1.87 |
| Theia_862 | HSC_1253 | 22 | 14.3 | 62.4 | 25.3 | 13.9 | 0.88 |
| Theia_345 | HSC_1112 | 22 | 15.8 | 24.6 | 17.9 | 18.7 | 1.35 |
| UBC_294 | CWNU_1362 | 22 | 17.5 | 12.7 | 8.7 | 70.6 | 1.0 |
| Theia_229 | HSC_782 | 22 | 11.8 | 62.2 | 20.8 | 25.4 | 0.79 |
| Haffner_13 | CWNU_2407 | 22 | 7.0 | 58.6 | 14.2 | 51.2 | 0.92 |
| Theia_863 | Theia_5 | 22 | 11.8 | 37.9 | 54.5 | 22.0 | 1.28 |
| HSC_2247 | HSC_453 | 22 | 15.3 | 32.2 | 69.7 | 24.7 | 1.25 |
| Theia_874 | OCSN_70 | 22 | 15.7 | 45.5 | 3.4 | 24.1 | 1.72 |
| Theia_1188 | ASCC_111 | 22 | 4.8 | 26.4 | 26.8 | 29.9 | 0.86 |
| UPK_492 | CWNU_331 | 22 | 15.7 | 21.4 | 7.8 | 23.9 | 1.5 |
| HSC_2315 | FSR_1452 | 22 | 11.9 | 24.0 | 23.3 | 28.3 | 0.81 |
| Theia_1232 | HSC_2718 | 22 | 8.8 | 14.5 | 81.2 | 24.6 | 0.88 |

| | | | Table 1. Continued | | | | |
|---|---|---|---|---|---|---|---|
| Name 1 | Name 2 | t | Min_dist, пк | $R_1$, пк | $R_2$, пк | $V_{rel}$, км/с | dlogAge |
| Theia_1528 | HSC_1640 | 22 | 18.5 | 11.6 | 38.9 | 22.4 | 1.64 |
| Theia_8014 | HSC_899 | 22 | 11.6 | 24.1 | 9.1 | 18.3 | 1.7 |
| HSC_691 | BDSB_30 | 22 | 12.7 | 17.5 | 8.8 | 17.8 | 1.23 |
| HSC_2881 | Bochum_13 | 22 | 4.6 | 5.6 | 15.4 | 22.5 | 1.23 |
| Theia_874 | OCSN_82 | 24 | 16.7 | 45.5 | 35.5 | 28.6 | 1.13 |
| NGC_6882 | HSC_544 | 24 | 16.3 | 40.2 | 16.2 | 6.4 | 1.07 |
| HSC_381 | FSR_0416 | 24 | 13.1 | 37.9 | 4.4 | 30.3 | 0.81 |
| UPK_317 | UBC_51 | 24 | 10.7 | 17.4 | 11.6 | 15.6 | 1.15 |
| HSC_1941 | CWNU_1032 | 24 | 12.6 | 12.9 | 37.9 | 28.9 | 0.81 |
| Theia_401 | HSC_835 | 24 | 5.1 | 72.2 | 13.3 | 17.0 | 1.47 |
| Theia_93 | HSC_1840 | 24 | 15.9 | 22.9 | 22.0 | 9.6 | 1.16 |
| HSC_1640 | HSC_725 | 24 | 19.3 | 38.9 | 13.9 | 17.6 | 1.66 |
| OCSN_28 | HSC_396 | 24 | 17.1 | 27.1 | 67.1 | 16.1 | 0.87 |
| Trumpler_3 | ASCC_125 | 24 | 5.9 | 64.4 | 4.2 | 17.5 | 1.02 |
| UPK_45 | HSC_419 | 24 | 17.6 | 24.6 | 18.8 | 32.5 | 0.95 |
| HSC_1692 | CWNU_1096 | 24 | 11.8 | 16.4 | 9.2 | 12.4 | 1.1 |
| Theia_1890 | HSC_1867 | 24 | 16.1 | 17.3 | 9.8 | 22.2 | 0.88 |
| Theia_1528 | HSC_106 | 24 | 2.9 | 11.6 | 5.9 | 50.8 | 2.1 |
| Theia_1408 | CWNU_172 | 24 | 13.3 | 18.5 | 14.0 | 23.4 | 0.78 |
| Mon_OB1-D | CWNU_410 | 24 | 14.8 | 3.8 | 22.1 | 15.7 | 1.12 |
| OCSN_70 | HSC_1169 | 24 | 15.3 | 3.4 | 22.2 | 24.1 | 1.03 |
| HSC_1973 | Alessi_170 | 24 | 16.6 | 13.5 | 39.6 | 34.1 | 1.23 |
| UBC_169 | Theia_420 | 24 | 12.4 | 10.8 | 26.2 | 26.6 | 1.07 |
| HSC_2796 | ASCC_87 | 26 | 16.2 | 16.0 | 16.8 | 39.2 | 1.6 |
| HSC_2144 | Gulliver_40 | 26 | 18.9 | 13.2 | 10.6 | 32.0 | 0.96 |
| Majaess_90 | CWNU_212 | 26 | 18.1 | 11.8 | 15.1 | 14.2 | 1.2 |
| OC_0395 | HSC_1693 | 26 | 9.4 | 21.2 | 13.7 | 14.1 | 0.86 |
| OCSN_54 | HSC_1112 | 26 | 14.2 | 18.9 | 17.9 | 14.6 | 1.82 |
| HSC_2468 | HSC_1991 | 26 | 18.7 | 24.3 | 15.7 | 6.4 | 1.2 |

Table 1. Continued

| Name 1 | Name 2 | t | Min_dist, пк | $R_1$, пк | $R_2$, пк | $V_{rel}$, км/с | dlogAge |
|---|---|---|---|---|---|---|---|
| Theia_704 | HSC_2437 | 26 | 19.4 | 30.5 | 30.6 | 11.5 | 1.07 |
| HSC_460 | CWNU_446 | 26 | 11.9 | 15.4 | 8.2 | 36.4 | 2.07 |
| UPK_120 | ASCC_125 | 26 | 19.7 | 24.0 | 4.2 | 14.7 | 1.24 |
| HSC_2202 | HSC_1958 | 26 | 19.6 | 10.2 | 13.6 | 21.0 | 1.72 |
| Theia_280 | HSC_789 | 26 | 16.5 | 52.3 | 11.3 | 13.6 | 1.54 |
| HSC_1894 | CWNU_1087 | 26 | 18.6 | 15.6 | 9.4 | 16.9 | 0.9 |
| HSC_1067 | HSC_824 | 26 | 16.8 | 44.0 | 4.8 | 17.3 | 1.76 |
| Theia_648 | NGC_2068 | 26 | 19.9 | 19.6 | 12.4 | 11.0 | 1.35 |
| UPK_126 | CWNU_209 | 26 | 16.1 | 5.6 | 11.3 | 24.1 | 1.6 |
| Theia_1890 | NGC_2539 | 26 | 16.4 | 17.3 | 16.4 | 28.2 | 1.62 |
| OC_0018 | HSC_2231 | 26 | 9.8 | 5.8 | 11.6 | 67.9 | 2.26 |
| Teutsch_182 | HSC_883 | 26 | 18.3 | 17.3 | 23.2 | 9.0 | 1.11 |
| HSC_957 | ASCC_125 | 26 | 12.2 | 18.0 | 4.2 | 16.9 | 1.62 |
| OCSN_91 | HSC_2850 | 26 | 14.0 | 4.9 | 36.5 | 8.7 | 1.36 |
| HSC_507 | HSC_207 | 26 | 14.2 | 28.9 | 15.1 | 13.6 | 0.96 |
| HSC_2785 | HSC_283 | 26 | 16.5 | 11.2 | 24.0 | 40.0 | 1.08 |
| Theia_345 | Teutsch_5 | 26 | 11.3 | 24.6 | 10.3 | 62.3 | 0.98 |
| UBC_1592 | HSC_2638 | 26 | 19.8 | 9.6 | 13.6 | 24.5 | 0.75 |
| HSC_703 | CWNU_1240 | 26 | 17.5 | 37.3 | 11.0 | 32.3 | 1.06 |
| HSC_272 | CWNU_1292 | 26 | 19.9 | 18.7 | 14.1 | 30.7 | 1.09 |
| UPK_540 | HSC_229 | 26 | 10.8 | 26.4 | 18.3 | 24.2 | 1.95 |
| HSC_1908 | CWNU_232 | 26 | 8.2 | 15.4 | 31.2 | 19.6 | 1.33 |
| HSC_2693 | HSC_2139 | 26 | 13.3 | 17.4 | 24.1 | 50.2 | 1.73 |
| UBC_302 | Collinder_272 | 26 | 19.6 | 8.6 | 19.0 | 40.5 | 0.91 |
| Theia_403 | HSC_2283 | 26 | 19.7 | 18.0 | 67.4 | 20.4 | 0.77 |
| Theia_366 | Ruprecht_26 | 26 | 10.2 | 19.6 | 15.6 | 10.4 | 0.75 |
| HSC_2636 | HSC_883 | 26 | 19.8 | 19.0 | 23.2 | 22.6 | 0.95 |
| Theia_3796 | OC_0497 | 28 | 18.3 | 5.5 | 5.6 | 51.4 | 1.45 |
| Theia_3715 | CWNU_164 | 28 | 5.1 | 13.6 | 13.5 | 38.5 | 1.31 |
| Table 1. Continued ||||||||
| Name 1 | Name 2 | t | Min_dist, пк | $R_1$, пк | $R_2$, пк | $V_{rel}$, км/с | dlogAge |
| HSC_725 | CWNU_168 | 28 | 17.6 | 13.9 | 13.9 | 30.7 | 1.53 |
| CWNU_2419 | CWNU_207 | 28 | 19.5 | 40.0 | 9.7 | 61.7 | 1.09 |
| HSC_926 | Collinder_135 | 28 | 19.8 | 35.1 | 14.1 | 21.1 | 0.85 |
| Theia_6 | OC_0257 | 28 | 15.9 | 18.8 | 11.9 | 11.5 | 1.25 |
| Theia_6599 | CWNU_1111 | 28 | 14.6 | 34.3 | 14.3 | 41.0 | 0.75 |
| OCSN_28 | HSC_598 | 28 | 19.2 | 27.1 | 33.1 | 13.9 | 0.96 |
| HSC_926 | Alessi_36 | 28 | 19.6 | 35.1 | 9.8 | 19.8 | 0.87 |
| HSC_1234 | HSC_1040 | 28 | 18.3 | 17.3 | 12.2 | 9.6 | 1.9 |
| UPK_614 | UFMG_66 | 28 | 16.5 | 13.7 | 11.8 | 37.6 | 0.95 |
| UPK_606 | Mamajek_2 | 28 | 11.6 | 14.6 | 45.4 | 6.9 | 1.01 |
| HSC_894 | HSC_627 | 28 | 19.3 | 17.0 | 4.9 | 30.6 | 1.88 |
| HSC_2022 | HSC_1913 | 28 | 19.0 | 7.2 | 8.2 | 18.1 | 1.91 |
| Theia_3980 | CWNU_406 | 28 | 14.5 | 20.5 | 15.7 | 15.1 | 1.21 |
| HSC_1122 | HSC_926 | 28 | 18.7 | 45.7 | 35.1 | 71.3 | 1.09 |
| UBC_182 | NGC_7082 | 28 | 19.3 | 15.2 | 14.2 | 28.3 | 0.8 |
| OCSN_28 | ASCC_101 | 28 | 4.0 | 27.1 | 47.9 | 9.8 | 1.02 |
| UPK_72 | HSC_1385 | 28 | 10.4 | 31.5 | 22.0 | 37.2 | 1.11 |
| OCSN_41 | CWNU_1216 | 28 | 18.3 | 7.0 | 29.5 | 7.7 | 1.42 |
| OC_0339 | COIN-Gaia_12 | 28 | 10.3 | 4.9 | 27.9 | 25.9 | 0.83 |
| UPK_470 | OC_0289 | 28 | 18.6 | 41.1 | 22.8 | 43.5 | 1.14 |
| FSR_1527 | CWNU_246 | 28 | 13.2 | 22.2 | 36.3 | 56.0 | 1.71 |
| Theia_246 | Alessi_36 | 28 | 16.0 | 31.5 | 9.8 | 10.0 | 0.82 |
| UBC_1622 | UBC_482 | 28 | 19.1 | 14.3 | 27.8 | 27.5 | 1.15 |
| Teutsch_5 | HSC_1523 | 28 | 8.5 | 10.3 | 10.3 | 64.0 | 1.89 |
| Theia_85 | ASCC_125 | 28 | 18.8 | 24.5 | 4.2 | 18.0 | 1.2 |
| HSC_1667 | HSC_1347 | 28 | 14.9 | 13.8 | 42.7 | 6.3 | 1.33 |
| HSC_2068 | HSC_453 | 28 | 12.0 | 15.8 | 69.7 | 8.8 | 0.97 |
| OC_0158 | HSC_620 | 28 | 16.8 | 6.7 | 10.7 | 21.9 | 1.14 |
| UBC_1295 | BDSB_93 | 28 | 17.6 | 20.6 | 17.2 | 39.1 | 1.22 |

Table 1. Continued

| Name 1 | Name 2 | t | Min_dist, пк | $R_1$, пк | $R_2$, пк | $V_{rel}$, км/с | dlogAge |
|---|---|---|---|---|---|---|---|
| HSC_1052 | HSC_465 | 28 | 18.6 | 7.9 | 13.1 | 36.2 | 1.35 |
| HSC_204 | CWNU_236 | 28 | 19.6 | 22.1 | 12.9 | 6.4 | 0.94 |
| UBC_374 | UBC_373 | 28 | 17.2 | 15.5 | 12.8 | 12.6 | 1.24 |
| HSC_2587 | HSC_2461 | 28 | 8.4 | 10.6 | 11.1 | 20.3 | 0.96 |
| Theia_665 | HSC_1958 | 30 | 17.5 | 22.1 | 13.6 | 19.5 | 1.49 |
| Theia_6680 | HSC_190 | 30 | 5.6 | 23.1 | 15.3 | 24.8 | 0.97 |
| Stock_8 | Ruprecht_145 | 30 | 19.6 | 21.7 | 65.7 | 81.3 | 1.77 |
| UPK_562 | BH_56 | 30 | 10.0 | 13.8 | 12.4 | 11.9 | 1.82 |
| OCSN_70 | HSC_977 | 30 | 19.8 | 3.4 | 23.4 | 29.8 | 1.13 |
| NGC_1960 | HSC_142 | 30 | 13.2 | 23.2 | 6.6 | 89.1 | 0.75 |
| UBC_669 | ESO_226-06 | 30 | 14.4 | 7.7 | 11.6 | 13.9 | 0.76 |
| UBC_1004 | CWNU_1920 | 30 | 19.8 | 16.8 | 29.5 | 32.4 | 0.8 |
| UPK_639 | CWNU_2145 | 30 | 15.6 | 70.9 | 20.3 | 70.8 | 1.08 |
| Theia_4877 | Gulliver_24 | 30 | 9.7 | 22.1 | 11.3 | 35.1 | 0.76 |
| Theia_6 | HSC_1194 | 30 | 17.4 | 18.8 | 11.0 | 14.2 | 1.28 |
| Theia_361 | HSC_1314 | 30 | 12.6 | 6.9 | 73.6 | 21.6 | 1.55 |
| Theia_3397 | Theia_1424 | 30 | 19.9 | 68.6 | 29.5 | 47.0 | 0.82 |
| Theia_72 | CWNU_1134 | 30 | 8.8 | 31.2 | 25.3 | 7.1 | 0.84 |
| IC_2391 | HSC_1378 | 30 | 18.6 | 31.5 | 16.5 | 32.4 | 0.87 |
| HSC_2603 | HSC_1724 | 30 | 18.3 | 22.5 | 21.7 | 9.7 | 1.37 |
| Theia_361 | CWNU_1129 | 30 | 13.6 | 6.9 | 21.7 | 23.1 | 2.44 |
| UBC_255 | OC_0322 | 30 | 14.6 | 21.0 | 13.9 | 31.5 | 2.26 |
| HSC_1314 | CWNU_1129 | 30 | 18.7 | 73.6 | 21.7 | 5.7 | 0.89 |
| HXHWL_7 | ASCC_90 | 30 | 13.3 | 7.5 | 60.2 | 26.0 | 0.78 |
| UPK_127 | UBC_10b | 30 | 14.2 | 8.1 | 13.5 | 12.1 | 1.14 |
| UPK_535 | HSC_1774 | 30 | 15.7 | 16.1 | 27.5 | 22.8 | 0.8 |
| HSC_2784 | HSC_2458 | 30 | 19.1 | 28.0 | 8.9 | 12.0 | 0.76 |
| PHOC_24 | Majaess_88 | 30 | 18.6 | 15.3 | 12.9 | 31.8 | 0.78 |
| Theia_1038 | OCSN_3 | 30 | 20.0 | 28.3 | 12.6 | 11.7 | 0.96 |

Table 1. Continued

| Name 1 | Name 2 | t | Min_dist, пк | $R_1$, пк | $R_2$, пк | $V_{rel}$, км/с | dlogAge |
|---|---|---|---|---|---|---|---|
| UPK_172 | Theia_1140 | 30 | 17.6 | 21.9 | 17.8 | 26.7 | 0.93 |
| UPK_185 | HSC_747 | 30 | 9.8 | 23.4 | 10.0 | 11.1 | 1.47 |
| CWNU_1801 | CWNU_1227 | 30 | 13.9 | 19.4 | 29.0 | 32.4 | 1.8 |
| OC_0470 | HSC_1426 | 30 | 18.8 | 18.4 | 28.3 | 13.4 | 1.06 |
| HSC_1900 | HSC_453 | 30 | 19.6 | 15.7 | 69.7 | 7.5 | 0.96 |
| OCSN_27 | Melotte_111 | 30 | 15.2 | 38.6 | 61.4 | 18.1 | 1.64 |
| Theia_246 | Collinder_135 | 30 | 11.8 | 31.5 | 14.1 | 9.5 | 0.81 |
| UPK_422 | Teutsch_5 | 30 | 19.1 | 46.7 | 10.3 | 59.7 | 1.83 |
| HSC_2503 | Gulliver_9 | 30 | 18.6 | 18.0 | 28.4 | 25.4 | 1.47 |
| HSC_606 | CWNU_58 | 30 | 15.8 | 25.2 | 12.3 | 21.0 | 1.09 |
| HSC_894 | ASCC_114 | 30 | 16.5 | 17.0 | 11.8 | 25.3 | 1.26 |
| HSC_974 | HSC_932 | 32 | 12.3 | 26.4 | 15.0 | 9.8 | 0.73 |
| Theia_387 | NGC_752 | 32 | 13.2 | 72.9 | 82.9 | 14.5 | 1.11 |
| IC_1396 | HXHWL_8 | 32 | 17.0 | 14.7 | 30.8 | 16.8 | 1.55 |
| HSC_2662 | HSC_558 | 32 | 11.2 | 25.1 | 19.8 | 16.8 | 1.31 |
| UPK_169 | UPK_53 | 32 | 19.4 | 20.0 | 17.1 | 18.8 | 1.43 |
| HSC_1827 | HSC_1658 | 32 | 19.5 | 10.6 | 25.3 | 3.7 | 0.89 |
| UBC_1456 | Loden_46 | 32 | 17.5 | 14.8 | 18.2 | 33.4 | 0.99 |
| HSC_618 | HSC_343 | 32 | 12.4 | 10.1 | 19.7 | 20.8 | 1.36 |
| Theia_713 | Riddle_6 | 32 | 11.9 | 94.3 | 6.2 | 17.5 | 1.54 |
| Harvard_13 | HSC_2719 | 32 | 13.5 | 27.6 | 22.1 | 32.6 | 0.98 |
| UPK_369 | HSC_2453 | 32 | 15.5 | 14.8 | 85.9 | 24.4 | 0.77 |
| Theia_6599 | NGC_2232 | 32 | 12.0 | 34.3 | 17.6 | 40.0 | 0.96 |
| Theia_4877 | HSC_862 | 32 | 16.7 | 22.1 | 8.2 | 29.5 | 2.07 |
| HSC_2022 | CWNU_384 | 32 | 10.9 | 7.2 | 15.5 | 7.8 | 0.97 |
| Theia_66 | Kronberger_92 | 32 | 17.0 | 11.7 | 13.2 | 33.0 | 1.42 |
| HSC_2971 | HSC_2836 | 32 | 19.8 | 41.0 | 18.8 | 30.9 | 1.05 |
| Theia_66 | OC_0370 | 32 | 18.6 | 11.7 | 10.4 | 26.5 | 1.41 |
| HSC_1516 | CWNU_1129 | 32 | 17.2 | 13.8 | 21.7 | 24.2 | 0.89 |

Table 1. Continued

| Name 1 | Name 2 | t | Min_dist, пк | $R_1$, пк | $R_2$, пк | $V_{rel}$, км/с | dlogAge |
|---|---|---|---|---|---|---|---|
| Theia_71 | ASCC_99 | 32 | 11.9 | 8.1 | 40.4 | 14.1 | 0.76 |
| OCSN_89 | HSC_453 | 32 | 13.6 | 34.9 | 69.7 | 15.2 | 0.82 |
| CWNU_338 | Alessi_170 | 32 | 11.6 | 36.1 | 39.6 | 31.7 | 1.72 |
| NGC_6793 | FSR_0551 | 32 | 18.3 | 47.4 | 14.4 | 25.7 | 1.32 |